\documentclass[9pt,twocolumn,twoside]{pnas-new-custom}
\usepackage{xspace}
\usepackage{ulem}

\newcommand\southr{\bgroup\markoverwith{\textcolor{orange}{\rule[0.5ex]{2pt}{0.4pt}}}\ULon}

\newcommand{\reb}{{\textsc \tt REBOUND}\xspace}
\newcommand{\whfast}{{\textsc \tt WHFast}\xspace}

\templatetype{pnasresearcharticle} 

\title{Predicting the long-term stability of compact multiplanet systems}

\author[a,1,2]{Daniel Tamayo}
\author[a]{Miles Cranmer} 
\author[b]{Samuel Hadden}
\author[c,d]{Hanno Rein}
\author[e]{Peter Battaglia}
\author[d,f]{Alysa Obertas}
\author[g,h]{Philip J. Armitage}
\author[h,a,i]{Shirley Ho}
\author[h]{David Spergel}
\author[j]{Christian Gilbertson}
\author[d]{Naireen Hussain}
\author[j,c,d]{Ari Silburt}
\author[k]{Daniel Jontof-Hutter}
\author[c,d,l]{Kristen Menou}

\affil[a]{Department of Astrophysical Sciences, Princeton University, Princeton, New Jersey 08544, USA}
\affil[b]{Center for Astrophysics | Harvard \& Smithsonian, 60 Garden St., MS 51, Cambridge, MA 02138, USA} 
\affil[c]{Department of Physical and Environmental Sciences, University of Toronto at Scarborough, Toronto, Ontario M1C 1A4, Canada}
\affil[d]{David A. Dunlap Department of Astronomy and Astrophysics, University of Toronto, Toronto, Ontario, M5S 3H4, Canada}
\affil[e]{Google DeepMind, London, UK}
\affil[f]{Canadian Institute for Theoretical Astrophysics, University of Toronto, Toronto, Ontario, M5S 3H8, Canada}
\affil[g]{Department of Physics and Astronomy, Stony Brook University, Stony Brook, NY 11790, USA}
\affil[h]{Center for Computational Astrophysics, Flatiron Institute, New York, NY 10010, USA}
\affil[i]{Department of Physics, Carnegie Mellon University, Pittsburgh, PA 15217, USA}
\affil[j]{Department of Astronomy and Astrophysics, The Pennsylvania State University, University Park, PA 16802, USA}
\affil[k]{Department of Physics, University of the Pacific, Stockton, CA 95211, USA}
\affil[l]{Department of Physics, University of Toronto, 60 St George Street, Toronto, Ontario, M5S 1A7, Canada}
\leadauthor{Tamayo} 

\significancestatement{
Observations of planets beyond our solar system (exoplanets) yield uncertain orbital parameters.
Particularly in compact multi-planet systems, a significant fraction of observationally inferred orbital configurations can lead to planetary collisions on timescales short compared to the age of the system.
Rejection of these unphysical solutions can thus sharpen our view of exoplanetary orbital architectures.
Long-term stability determination is currently performed through direct orbital integrations.
However, this approach is computationally prohibitive for application to the full exoplanet sample.
By speeding up this process by up to five orders of magnitude, we enable precise exoplanet characterization of compact multi-planet systems and our ability to examine the stability properties of the multi-planet exoplanet sample as a whole.}

\authorcontributions{\textsuperscript{1} NHFP Sagan Fellow}
\correspondingauthor{\textsuperscript{2}To whom correspondence should be addressed. E-mail: dtamayo@astro.princeton.edu}


\begin{abstract}
We combine analytical understanding of resonant dynamics in two-planet systems with machine learning techniques to train a model capable of robustly classifying stability in compact multi-planet systems over long timescales of $10^9$ orbits.  
Our Stability of Planetary Orbital Configurations Klassifier (SPOCK) predicts stability using physically motivated summary statistics measured in integrations of the first $10^4$ orbits, thus achieving speed-ups of up to $10^5$ over full simulations.
This computationally opens up the stability constrained characterization of multi-planet systems.
Our model, trained on $\approx 100,000$ three-planet systems sampled at discrete resonances, generalizes both to a sample spanning a continuous period-ratio range, as well as to a large five-planet sample with qualitatively different configurations to our training dataset.  
Our approach significantly outperforms previous methods based on systems’ angular momentum deficit, chaos indicators, and parametrized fits to numerical integrations.
We use SPOCK to constrain the free eccentricities between the inner and outer pairs of planets in the Kepler-431 system of three approximately Earth-sized planets to both be below 0.05.
Our stability analysis provides significantly stronger eccentricity constraints than currently achievable through either radial velocity or transit duration measurements for small planets, and within a factor of a few of systems that exhibit transit timing variations (TTVs).
Given that current exoplanet detection strategies now rarely allow for strong TTV constraints (Hadden et al., 2019), SPOCK enables a powerful complementary method for precisely characterizing compact multi-planet systems.
We publicly release SPOCK for community use.
\end{abstract}

\dates{Received for review January 22, 2020. Accepted June 15th, 2020. Published week of July 13th, 2020. arXiv compiled Jul 13, 2020. NOT ACTUAL PNAS ARTICLE}
\doi{Published version available at: \url{http://www.pnas.org/cgi/doi/10.1073/pnas.2001258117}}

\begin{document}

\maketitle
\thispagestyle{firststyle}
\ifthenelse{\boolean{shortarticle}}{\ifthenelse{\boolean{singlecolumn}}{\abscontentformatted}{\abscontent}}{}

\dropcap{I}saac Newton, having formulated his law of gravitation, recognized that it left the long term stability of the Solar System in doubt. Would the small near-periodic perturbations the planets exert on one another average out over long timescales, or would they accumulate until orbits cross, rendering the system unstable to planetary collisions or ejections? 

The central difficulty arises from the existence of resonances, where there is an integer ratio commensurability between different frequencies in the system. 
These resonances complicate efforts to average the dynamics over long timescales. Work on this problem culminated in the celebrated Komolgorov-Arnold-Moser (KAM) theorem \citep{Kolmogorov54, Moser62, Arnold63}, which guarantees the existence of stable, quasiperiodic trajectories below a specified perturbation strength. 
Unfortunately, the KAM theorem is generally not informative in this context, since it typically can only guarantee stability for masses far below the planetary regime \citep{Henon66, Celletti05}. 

Without a clear path to a full solution, we focus on the limit of closely separated planets.
This regime has important applications for understanding the orbital architectures of planetary systems beyond our own (exoplanets), since strong observational biases toward detecting planets close to their host star result in compact populations of observed multiplanet systems\footnote{We henceforth define compact 3+ planet systems as having at least one trio of planets with period ratios between adjacent planets < 2. This currently represents $\approx 40$\% of all observed 3+ planet systems.}.
In these dynamically delicate configurations, it is possible for most of the orbital solutions inferred from noisy data to undergo violent dynamical instabilities when numerically integrated forward in time for even $0.1\%$ of the system's age \citep{Mills16, Gillon17}.
Since one does not expect to discover most systems just prior to such a cataclysm, this offers an opportunity to constrain the masses and orbital parameters of such planets by rejecting configurations that lead to rapid instability.
In this way, previous authors have performed direct numerical (N-body) integrations to narrow down physical orbital architectures and formation histories for important exoplanet discoveries \citep[e.g.,][]{Steffen13, Tamayo15, Wang18, Quarles17, Tamayo17}.

However, given the high dimensionality of parameters, the computational expense of such long-term integrations typically results in only a small fraction of candidate orbital configurations being explored, and integration timespans being many orders of magnitude shorter than the typical Gyr ages of such systems \citep[e.g.,][]{Rivera10, Jontof14, Buchhave16, Mills16, Tamayo17, Quarles17, Hadden17, Grimm18}. 
This renders the widespread application of such constraints to the ever-growing exoplanet sample orders of magnitude beyond computational reach.

Extensive previous work has narrowed down the particular resonances responsible for dynamical instabilities in compact systems.
In particular, analytical studies of tightly spaced {\it two}-planet systems \citep{Wisdom80, Deck13, Hadden18} have shown that the chaos is driven specifically by the interactions between mean motion resonances (MMRs), i.e., integer commensurabilities between planets' orbital periods.
The limited number of available MMRs in two-planet systems implies that for initially circular orbits, there exists a critical, mass-dependent separation between the two bodies.
For planetary separations below this limit, MMRs are close enough to one another in phase space to overlap and drive rapid instabilities \citep{Wisdom80, Deck13}, and there is a sharp transition to long-lived configurations beyond it.
This result has recently been generalized for eccentric orbits \citep{Hadden18}.

By contrast in 3+ planet systems, instabilities can occur for separations between adjacent planet pairs significantly beyond the above two-planet limit, and instability times exhibit a continuous and much larger dynamic range \citep{Chambers96}. 
Previous work has argued that this can not be solely explained by the larger number of available MMRs between all possible pairs of planets \citep{Quillen11, Quillen14}.
These authors argue that three-body resonances, i.e., integer combinations between the periods of three bodies are responsible for ``filling in the space" between two-body MMRs and driving instabilities over a continuous range of separations. 

But while a clearer physical picture is emerging, theoretical estimates can not yet quantitatively match the results from numerical integrations \citep{Quillen11}.
Many previous numerical studies have instead presented empirical fits to the overall steep rise in instability times with interplanetary separation, recorded from large suites of numerical integrations \citep{Chambers96, Yoshinaga99, Marzari02, Zhou07, Faber07, Smith09, Matsumoto12, Pu15}.
This is a useful approach for elucidating the underlying dynamics and scalings with dominant parameters, but typically involves simplifications such as equal-mass, or equal-separation planets.
This limitation, together with modulations near MMRs on overall trends in instability times of up to five orders of magnitude \citep{Obertas17}, lead to quantitative disagreements between such studies, and render them inadequate for accurately characterizing real multiplanet systems (see Sec.\:\ref{results}).

Here, we present a machine learning model that can reliably classify the stability of compact 3+ planet configurations over $10^9$ orbits.
Our model, the Stability of Planetary Orbital Configurations Klassifier (SPOCK), is up to $10^5$ times faster than direct integration, computationally opening up the stability constrained characterization of compact multi-planet systems.

\section{Previous Models} \label{prev}
Previous numerical efforts to predict the instability times of various orbital configurations can roughly be broken down into four groups:

\subsection{N-body} \label{nbody}
The most straightforward (and computationally costly) method is to run a direct numerical integration. 
A $10^9$ orbit integration with a timestep of $3.5\%$ of the innermost planet's orbital period takes $\approx 7$ CPU hours on a 2.1 GHz Intel Xeon Silver 4116 using the {\tt WHFast} integrator \citep{ReinTamayo15}. 

Interestingly, even this answer will not be perfect.
The fact that planetary systems are chaotic means that a given initial condition should not be considered to have a single instability time.
Rather, an N-body integration can be interpreted as sampling a single instability time from a broader distribution of values. 
If one numerically characterizes the distribution of these instability times, one finds that, for compact systems destabilizing within $10^9$ orbits, they are approximately log-normally distributed, with a uniform standard deviation of $\approx 0.4$ dex \citep{Hussain19, Rice18}.
To empirically quantify this fundamental limit to predictability, for each of the integrations in our training dataset, we have run a second ``shadow integration'' of the same initial conditions offset by one part in $10^{-11}$.
This represents an independent draw from that initial condition's instability time distribution.
There will thus be cases where one integration says the configuration is stable, while the other one does not.
The existence of these uncertain outcomes sets the fundamental limit any stability classifier can hope to reach.

\subsection{Hill}
Several previous studies have fit functional forms to instability times recorded in large suites of N-body integrations \citep[e.g.,][]{Chambers96, Marzari02, Faber07, Smith09, Obertas17}.
They found that instability times rise steeply with increasing interplanetary separation measured in mutual Hill radii, i.e., the characteristic radius around the planets in which their gravity dominates that of the star \citep[see also][]{Quillen11, Yalinewich19},
\begin{equation} \label{RH}
R_H = a_i \Bigg( \frac{m_i+m_{i+1}}{M_\star}\Bigg)^{1/3,}
\end{equation}
where $a_i$ is the semimajor axis of the inner planet in the pair, $m_i$ and $m_{i+1}$ are the respective planet masses, and $M_\star$ is the stellar mass\footnote{We note that the Hill sphere scales as the planet-star mass ratio $\mu$ to the one third power. Other authors \citep[e.g.,][]{Quillen11, Hadden18, Yalinewich19} argue that a $\mu^{1/4}$ scaling is better motivated. These scalings are close to one another, and given the poor performance of such models (Sec.\:\ref{results}) we do not pursue this possible correction.}.
While this provides insight into the underlying dynamics \citep{Quillen11, Yalinewich19}, other orbital parameters also strongly influence stability.
Follow-up studies have considered the effects of finite eccentricities and inclinations \citep[e.g.,][]{Yoshinaga99, Zhou07, Funk10, Wu19}, but make various simplifying assumptions (e.g., equal interplanetary separations and eccentricities).
Different assumptions lead to quantitative disagreements between different studies, and the reliability of their predictions to real systems, where all planets have independent orbital parameters, is unclear.

\subsection{AMD}
A classical result in orbital dynamics is that if the effects of MMRs are removed, then planets will exchange angular momenta at fixed semimajor axes \citep{Murray99}. 
Instabilities can still arise under these so-called secular dynamics, through chaos introduced by the overlap of resonances between the slower set of frequencies at which the orbits and their corresponding orbital planes precess \citep{Lithwick11secularchaos, Batygin15c}. 
In this approximation, there is a conserved quantity \citep{Laplace84,Laskar90}, termed the Angular Momentum Deficit (AMD).
The AMD acts as a constant reservoir of eccentricity and inclination that the planets can exchange among one another.
If the AMD is too small to allow for orbit crossing and collisions even in the worst case where {\it all} the eccentricity is given to one adjacent pair of planets, the system is AMD stable \citep{Laskar00, Laskar17}.
This is a powerful and simple analytic criterion, but it has two important caveats.
First, because it is a worst-case-scenario estimate, it yields no information on instability timescales for AMD {\it unstable systems}. 
For example, the Solar System is AMD unstable, but most integrations ($\approx 99\%$) of the Solar System nevertheless remain stable over the Sun's main sequence lifetime \citep{Laskar09}.
Second, the assumed secular model of the dynamics ignores the effects of MMRs, which for closely packed systems are typically nearby \citep[e.g.,][]{Migaszewski12}, and are an important source of dynamical chaos \citep[for a generalization of AMD stability in the presence of MMRs in the two-planet case, see][]{Petit17}.

\subsection{MEGNO}
Several authors have also used chaos indicators numerically measured from short integrations as a proxy for instability \citep{Migaszewski12, Marzari14}.
This is appealing given that systems that go unstable typically exhibit chaotic dynamics on shorter timescales.
A widely used chaos indicator is the Mean Exponential Growth factor of Nearby Orbits, or MEGNO \citep{Cincotta03}.
However, a planetary system can be chaotic yet never develop destructive instabilities on astrophysically relevant timescales.
Additionally, and most importantly, short integrations will fail to measure chaos on timescales longer than those simulated, potentially misclassifying systems that destabilize on long timescales.

\section{Predicting Long-Term Stability} \label{training}

Point-source Newtonian gravity is scale invariant. 
We exploit this fact by expressing all masses relative to that of the central star, and all times and distances in units of the innermost planet's orbital period and semimajor axis, respectively. 

Non-dimensionalizing timescales in this way is important when comparing systems with different absolute ages.
For example, the $\sim 40$ Myr age of the HR 8799 planets, with an innermost orbital period of $\approx 40$ yrs, only represents $10^6$ orbits \citep{Wang18}. 
For these short timescales, numerical integrations\footnote{Computation time scales linearly with the number of orbits and requires $\approx 10$s per million orbits with optimized algorithms \citep{Wisdom91} and current hardware.} are within reach, and SPOCK is not needed \citep{Wang18}. 

However, young multi-planet systems with long orbital periods are currently exceedingly rare in the exoplanet sample.
Population statistics and strong observational biases result in a multi-planet sample predominantly with innermost orbital periods of $\approx 0.01-0.1$ yrs, around stars that are several Gyr old.
We are thus most often interested in stability over timescales of $10^{11}-10^{12}$ orbits, which are computationally prohibitive for the number of candidate orbital configurations that typically require evaluation.

One approach would be to frame the task as a regression problem and predict an instability time for a given initial configuration.
However, given that most systems have large dynamical ages $>10^{11}$ orbits, for many applications one is simply interested in a binary classification between short-lived and long-term stable systems.
We therefore pursue a simpler binary classifier here, and defer a regression algorithm to future work (Cranmer et al., {\it in prep.}).

Any binary stability classification must specify a timescale, and we choose a value of $10^9$ orbits. 
Exoplanet systems may indeed continually undergo instabilities and restructurings over their lifetimes \citep[e.g.,][]{Volk15, Pu15}.
In such a situation, the quick removal of short-lived configurations naturally leaves systems with instability times comparable to their ages \citep{Laskar90}.
In that case, requiring stability over $10^{11}$ orbits in a comparably aged system could potentially throw out the true orbital configuration.
However, one could still reasonably reject configurations that destabilize within $<10^9$ orbits, given the low likelihood of finding a system within $<1\%$ of its lifetime of going unstable.
Following previous authors \citep{Smith09, Obertas17}, we define instability as when a pair of planets start crossing their sphere of gravitational influence (see {\it Materials and Methods}).

For the remainder of the paper we refer to configurations that survive $10^9$ orbits as {\it stable}, and ones that do not as {\it unstable}.

\subsection{Training SPOCK} \label{trainspock}

We frame our task as a supervised machine learning problem.
We begin by generating a large suite of $\approx 100,000$ initial conditions (described below), and perform the computationally expensive numerical integrations over $10^9$ orbits to empirically label each example as stable or unstable (see Sec.\:\ref{integrations}).
We take 80\% of these examples as a training set for our classifier, and use the remaining 20\% as a holdout set to test for potential overfitting with examples that were never encountered during training (Sec.\:\ref{holdout}).

\begin{figure*}
\centering
\includegraphics[width=.8\linewidth]{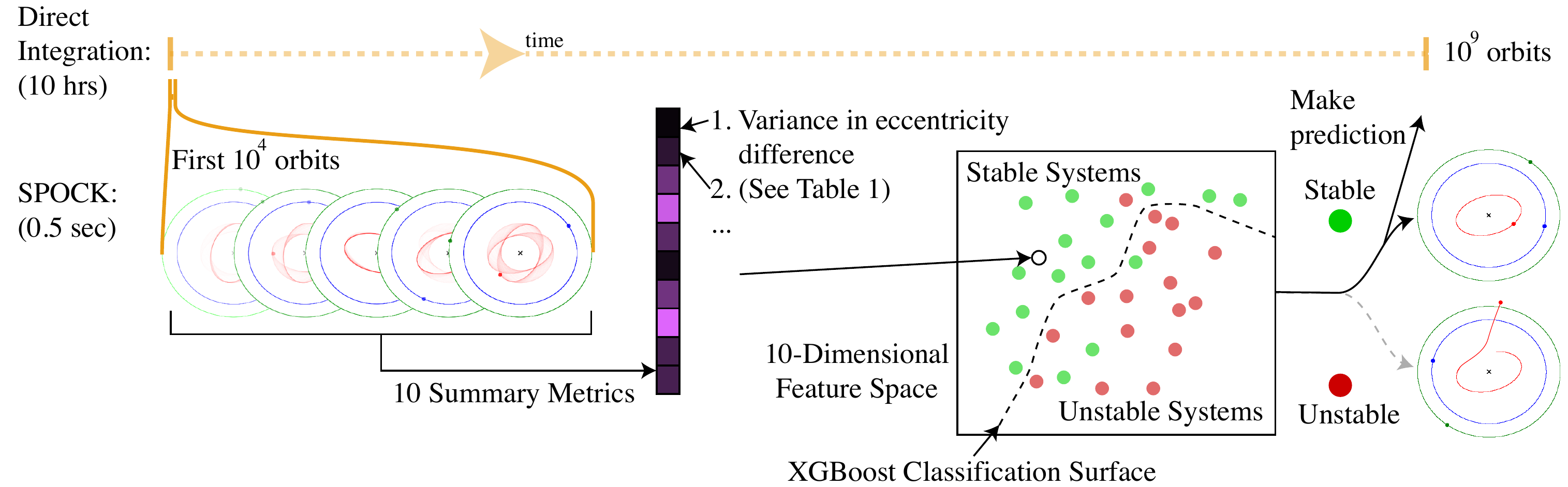}
\caption{A schematic illustrating how SPOCK classifies stability of a given initial configuration. 
The traditional approach numerically integrates the system for $10^9$ orbits, requiring roughly 10 hours with current hardware. 
SPOCK runs a much shorter $10^4$ orbit integration, and from it generates a set of 10 summary features (see Table \ref{summarytable}). 
These map to a point (white circle) in a 10-dimensional space, in which we have trained an {\tt XGBoost} model to classify stability.
SPOCK outputs an estimated probability that the given system is stable over $10^9$ orbits, up to $10^5$ times faster than direct integration.
}
\label{schematic}
\end{figure*}

The input to SPOCK is then a complete initial orbital configuration: stellar and planetary masses, along with 6 orbital elements or positions and velocities for each of the planets.
Our strategy for making a stability prediction is to first to run a computationally inexpensive integration of only $10^4$ orbits and, from this short snippet, numerically measure dynamically informative quantities (see Sec.\:\ref{features}).
Given that the machine learning model evaluation is effectively instantaneous, this represents a speedup factor of up to $10^5$.
This feature engineering step allows us to take a high dimensional set of inputs and reduce it to ten features that more compactly encode our partial understanding of the dynamics.
We then train a machine learning classifier to take this set of summary features as input to predict the probability that the system is stable over $10^9$ orbits. 
This is illustrated in Fig.\:\ref{schematic}.

Following a previous proof of concept \citep{Tamayo16}, we use the gradient-boosted decision tree algorithm {\tt XGBoost} \citep{Chen16}.
We found it significantly outperformed simple random forest and support vector machine implementations.
However, the specific choice of {\tt XGBoost} was not critical.
In an early comparison, we found similar results training a deep neural network (multi-layer perceptron) on the same features (see also \cite{Lam18} for an application to circumbinary planets).
The most important factor for performance was the adopted set of summary metrics. 
We are also exploring using deep learning to generate new features from the raw time series (Cranmer et al., \textit{in prep}).

\begin{figure}
\centering
\includegraphics[width=.7\linewidth]{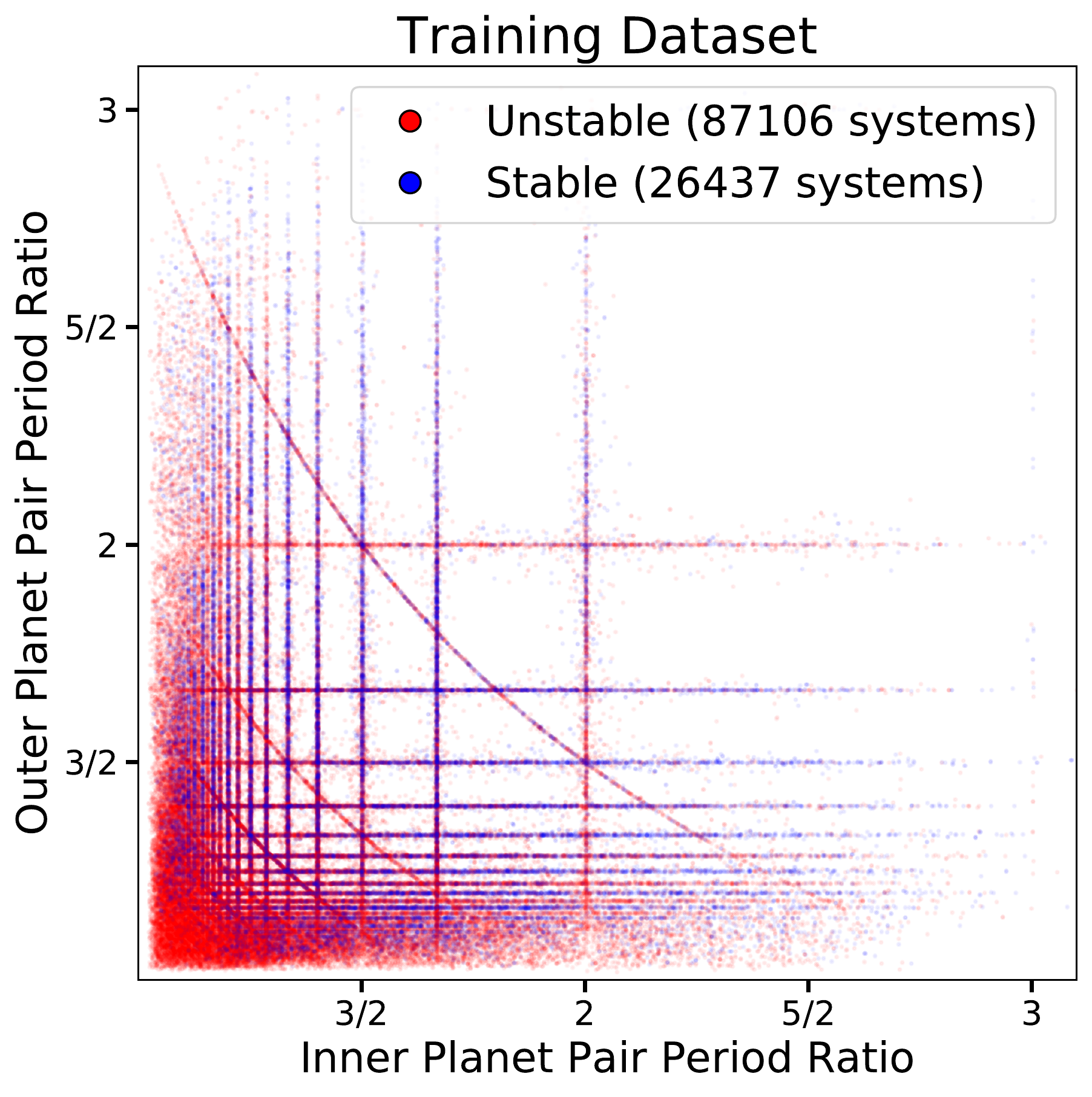}
\caption{Training dataset of three-planet systems in and near mean motion resonances (MMRs), which we posit drive instabilities in most compact multiplanet systems \citep{Wisdom80, Quillen11, Obertas17}. 
These cluster in lines at integer period ratios between adjacent planets.
The lifetimes of configurations within each of those lines can vary drastically depending on additional parameters (masses, eccentricities, phases) not visible in this projection, causing stable (blue) and unstable (red) systems to overlap.
\label{datasets}
}
\end{figure}

Going beyond exploratory machine learning models \citep{Tamayo16} to a robust classifier applicable to observed compact systems required several qualitative changes.
First, we relax their assumption of equal-mass planets, and we extend their horizon of stability prediction ($10^7$ orbits) by a factor of one hundred to a relevant timescale for real systems.
Second, previous work \citep{Wisdom80, Quillen11, Deck13} suggests that instabilities in compact systems are driven by the overlap of MMRs.
Rather than sampling phase space uniformly like \cite{Tamayo16}, we therefore choose to generate our training dataset of three-planet systems in and near such MMRs.
This fills in our training sample at the locations in phase space where the dynamical behavior is changing most rapidly, and we suspect this identification of the dominant dynamics is partially responsible for the excellent generalization to more general systems presented in Sec.\:\ref{uniform} and \ref{5p}.
Fig.\:\ref{datasets} shows our training set, plotting the period ratio between the inner two planets against that of the outer two planets.

Finally, the sharp changes in dynamics at each resonance (narrow lines visible in Fig.\:\ref{datasets}) make predictions challenging in the space of traditional orbital elements (orbital periods, eccentricities, etc.).
Indeed, we found that the model of \cite{Tamayo16} performed poorly near MMRs.
In this study, we exploit analytical transformations \citep{Sessin84, Hadden19} that isolate the effect of a particular resonance between a single pair of planets ({\it Materials and Methods}).
This allows us to both effectively sample the appropriate ranges in period ratios and other orbital elements (like eccentricities) not visible in the projection of Fig.\:\ref{datasets}, and to calculate several summary features in this smoother transformed space that makes predictions simpler than in the original sharply punctuated space of Fig.\:\ref{datasets} (see Sec.\:\ref{features}).

\subsection{Training set} \label{trainingset}

The two-planet case is analytically solvable \citep{Wisdom80, Deck13, Hadden18}, while for 3+ planet systems there is a qualitative change toward a continuous range of instability times over wider interplanetary separations \citep{Chambers96}.
We posit that instabilities driven by MMR overlap in higher multiplicity systems can be approximated by considering only adjacent planet trios.
Our training set thus consists only of compact three-planet systems, and we later test the trained model's generalization to higher multiplicity systems (Sec.\:\ref{5p}).
This is an important empirical test since, if true, it implies a robustness of our stability classifications to distant unseen planets.
This is crucial for reliable stability constrained characterization of exoplanet systems, and is consistent with previous numerical experiments with equal separation planets showing an insensitivity to additional bodies beyond somewhat larger multiplicities of five \citep{Chambers96}, as well as theoretical arguments showing that the Fourier amplitudes of the perturbation potential due to an additional planet fall off exponentially with separation \citep{Quillen11, Yalinewich19}.

To enable application to real systems, we sample unequal-mass unequal-separation, mutually inclined, eccentric initial orbital three-planet configurations for our training set, drawing from parameter ranges typically encountered in the current multiplanet sample.

In particular, the vast majority of 3+ planet systems have been discovered by the {\it Kepler} and {\it K2} missions as the bodies pass in front of (transit) their host star.

This implies that nearly all such systems must be approximately co-planar; otherwise the planets would not all cross in front of the star from our vantage point \citep[e.g.,][]{Fabrycky14}.
We therefore sample inclinations (log-uniformly and independently) from a narrow range of [$10^{-3},10^{-1}$] radians (where the upper limit has been extended somewhat beyond the mutual inclinations typically inferred to also allow the modeling of additional (unobserved) non-transiting planets).
The azimuthal orientations of the orbital planes (i.e., the longitudes of the ascending nodes) were drawn uniformly from $[0,2\pi]$.
This corresponds to maximum mutual orbital inclinations of $\approx 11^\circ$.

Most planets ($\approx 85\%$) in the current sample of compact 3+ planet systems, where stability constraints are most informative, are smaller than Neptune.
We therefore choose to independently and log-uniformly sample mass ratios to the central star from $10^{-4}$ ($\approx 2\times$ that of Neptune to the Sun) down below the typical threshold of detectability to $10^{-7}$ ($\approx$ 1/3 that of Mars to the Sun).

Any measure of dynamical compactness must incorporate these planetary masses.
This is often expressed in terms of the separations between adjacent planets in units of their mutual Hill radius (Eq.\:\ref{RH}).
We always initialize the innermost planet's semimajor axis at unity (since as mentioned above we work in units of the innermost semimajor axis), and choose to sample the separations between adjacent planets in the range from [0,30] $R_H$.
This encompasses $\approx 80$\% of the currently known planets in well characterized 3+ planet systems \citep{Weiss18}.
For scale, 30 $R_H$ also roughly corresponds to the wider dynamical separation between the terrestrial planets in our Solar System.

In particular, we randomly choose a planet pair (inner, outer or non-adjacent), and randomly sample their remaining orbital parameters in or near a randomly chosen MMR within 30 $R_H$ as described in detail in {\it Materials and Methods}.
Finally, we draw the remaining planet's separation from its nearest neighbor uniformly in the range [0,30] $R_H$.
This gives rise to the extended lines in Fig.\:\ref{datasets}.
Two of the planets are initialized at a particular resonant ratio (e.g., 3/2 on the x-axis), while the third planet's period can span a continuous range across different configurations and is not necessarily strongly influenced by MMRs.

Orbital eccentricities and phases for the resonant pair are described in {\it Materials and Methods}, while the third planet's orbital eccentricity is drawn log-uniformly between the characteristic eccentricities imparted when inner planets overtake their outer neighbors (approximated as the ratio of the interplanetary forces to the central force from the star), and the nominal value at which adjacent orbits would cross 
\begin{equation} \label{ecross}
e_{\rm cross} = (a_{i+1}-a_i)/a_{i+1}~.
\end{equation} 
Pericenter orientations and phases along the orbit for the remaining planet are drawn uniformly from $[0, 2\pi]$.
Finally, we reject any configurations that destabilize within $10^4$ orbits.
We refer to this dataset of 113,543 systems as the ``resonant dataset''.

\subsection{Machine Learning Model} \label{features}
We have a choice of what set of features about each system to pass the {\tt XGBoost} classifier.
For example, one could fully characterize a system by passing each planet's mass ratio with the star and its initial Cartesian positions and velocities. 
But presumably it would be easier for the algorithm if one transformed those initial conditions to more physical parameters, like the orbital elements (semimajor axis, eccentricity, etc.).
We instead choose to run a computationally inexpensive integration over $10^4$ orbits and numerically measure ten dynamically relevant quantities at 80 equally spaced outputs.

We experimented with different lengths of integrations and number of outputs.
Given an integration time series, our implementation takes a few tenths of a second to compute the summary features and evaluate the {\tt XGBoost} model.
This is about the time required to run $10^4$ orbits through N-body, so given this fixed overhead, there is no computational gain in running a shorter integration.
We found that the performance gain from longer integrations was marginal, but if the feature and model evaluations were optimized (e.g., ported to C), a more careful optimization of these parameters could be valuable.

For our first two features, we store the median of the MEGNO chaos indicator \citep{Cincotta03} (over the last 10\% of the short integration) and its standard deviation (over the last 80\% of the time series to avoid initial transients) as {\tt MEGNO} and {\tt MEGNOstd}, respectively.
We also record the initial values {\tt EMcross} of $e_{\rm cross}$ (Eq.\:\ref{ecross}) for each adjacent planet pair, and we use the smallest value to identify one adjacent pair of planets as `near' and the other as `far' for these and all remaining features.

The remaining six summary statistics capture the resonant dynamics.
In particular, in the near-resonant two-planet limit, only one particular combination of the eccentricities (approximately their vector difference, see {\it Materials and Methods}) matters \citep{Sessin84, Hadden19}, 
\begin{equation} \label{eminus}
{\bf e_-} \equiv {\bf e}_{i+1} - {\bf e}_{i},
\end{equation}
where ${\bf e}_i$ is a vector pointing toward the $i$-th planet's orbital pericenter, with a magnitude given by the orbital eccentricity.

Two of our summary features (one for each adjacent planet pair) are the standard deviation of $|{\bf e}_-|$ over the timespan of the short $10^4$ orbit integration, which we normalize through Eq.\:\ref{ecross} to the value required for that planet-pair to cross.
Qualitatively, this can help the classifier differentiate between configurations that oscillate close to a resonant equilibrium (small variations) and are dynamically protected by the MMR, versus configurations far from equilibrium where the MMR induces large-amplitude, often destabilizing, variations.

For each adjacent planet pair, we also search for the strongest j:j-k MMR within 3\% of the pair's period ratio, and record its non-dimensionalized strength, 
\begin{equation}
s = \sqrt{\frac{m_{i} + m_{i+1}}{M_\star}} \frac{(e_-/e_{\rm cross})^{k/2}}{(jn_{i+1} - (j-k)n_i)/n_i},
\end{equation}
where the $m_i$ are the planet masses, and the $n_i$ are the orbital mean motions ($n_i = 2\pi / P_i$ with $P_i$ as the orbital periods).
This is the appropriate expression when linearizing the dynamics, omitting a period-ratio-dependent prefactor that is comparable for all the nearby resonances \citep{Hadden19}. 
It is thus adequate for identifying the strongest nearby MMR, and we store its median value over the short integration as {\tt MMRstrengthnear} and {\tt MMRstrengthfar}.

Finally, we record the standard deviation of $|{\bf e}_+|$, a complementary combination of eccentricities to ${\bf e}_-$ that is approximately conserved \citep{Sessin84, Hadden19} in the single resonance, two-planet model (see {\it Materials and Methods}).
Providing SPOCK with the variation of this putatively conserved $|{\bf e}_+|$ variable quantifies the validity of our simple analytic model.
In particular, the analytical transformation is useful along isolated lines in Fig.\:\ref{datasets}, where a single resonance dominates the dynamics.
The transformation breaks down (and $|{\bf e}_+|$ can vary significantly) at locations where resonances cross and more than one resonance drives the dynamics, as well as in the blank space between resonances in Fig.\:\ref{datasets}.
However, these are typically also the easier regions to classify.
Line crossings are regions where resonances are typically strongly overlapped to drive rapid chaos \citep{Chirikov79}, and the dynamics vary more smoothly in the regions between strong resonances.
The complementarity and flexibility of these ten features allow SPOCK to reliably classify stability in a broad range of compact configurations.

We calculate these 10 features (summarized in Table \ref{summarytable}) for all initial conditions in our resonant dataset and then use them to train a gradient-boosted decision tree {\tt XGBoost} model \citep{Chen16}.
We adopt an 80\%-20\% train-test split, performing five-fold cross-validation on the training set.
We optimized hyperparameters to maximize the area under the ROC curve (AUC, Fig.\:\ref{summary}) using the {\tt hyperopt} package \citep{Bergstra13}. We provide our final hyperparameter values and ranges sampled in a {\tt jupyter} notebook in the accompanying repository, which trains the model.

In Table \ref{summarytable} we also list the relative feature importances in our final model, which measure the occurrence frequency of different features in the model's various decision trees.
All provide comparable information, partially by construction.
We started with a much wider set of 60 features, iteratively removing less important ones.
This marginally decreased the performance of our final classifier, but this is compensated by the improved interpretability of our simplified feature set.
While the feature importances are close enough that one should not overinterpret their relative values, it is clear that the resonant features are providing important dynamical information.

\begin{table}
\centering
\caption{Summary features in our trained model, ranked by their relative importance (see Sec.\:\ref{training}\ref{features} for discussion).
The smallest value of {\tt EMcross} is used to label one adjacent pair of planets as `near' and the other as `far'.
\label{summarytable}}
\begin{tabular}{lll} 
Feature Name & Description & Importance \\
\midrule
{\tt EMcrossnear} & Initial orbit-crossing $e_-$ value & 6844 \\
{\tt MMRstrengthnear} & Median strength of nearest MMR & 6568 \\
{\tt MMRstrengthfar} & Median strength of nearest MMR & 6392 \\
{\tt EPstdnear} & Stdev of $e_+$ mode & 6161 \\
{\tt EMfracstdfar} & Stdev of $e_-$ mode / EMcross & 5815 \\
{\tt EMfracstdnear} & Stdev of $e_-$ mode / EMcross & 5509 \\
{\tt EMcrossfar} & Initial orbit-crossing $e_-$ value & 5077 \\
{\tt EPstdfar} & Stdev of $e_+$ mode & 5009 \\
{\tt MEGNOstd} & Stdev of chaos indicator & 4763 \\
{\tt MEGNO} & Chaos indicator & 4350 \\
\bottomrule
\end{tabular}
\end{table}

\section{Results} \label{results}
\subsection{Holdout Set Performance} \label{holdout}

The accuracy of any classifier depends on the dataset. 
For example, it would be much harder to determine stability over $10^9$ orbits on a set of configurations right at the boundary of stability, which all went unstable between $10^8-10^{10}$ orbits, than on a dataset of configurations that either go unstable within $10^3$ or survived beyond $10^{15}$ orbits.
Thus, to avoid any straw-man comparisons to previous work, we follow a parallel process of training an {\tt XGBoost} model using the quantities (features) considered by previous authors.
This allows each model to optimize its thresholds for the training set at hand, providing a fair comparison.
In particular, for `N-body,' we ask the {\tt XGBoost} model to predict stability based on the instability time measured in the shadow integration (Sec.\:\ref{prev}\ref{nbody}).
For `MEGNO,' we measure the chaos indicator over the same short $10^4$ orbit integration as our model and pass this as a single feature to a separate {\tt XGBoost} model.
For `AMD,' we train on two features: the system's total AMD as a fraction of the critical AMD needed for a collision between each pair of adjacent planets \citep{Laskar17}.
Finally, for `Hill,' we train an {\tt XGBoost} model on the separations between adjacent planets as a fraction of their mutual Hill sphere radius (Eq.\:\ref{RH}).

\begin{figure}
\centering
\includegraphics[width=.7\linewidth]{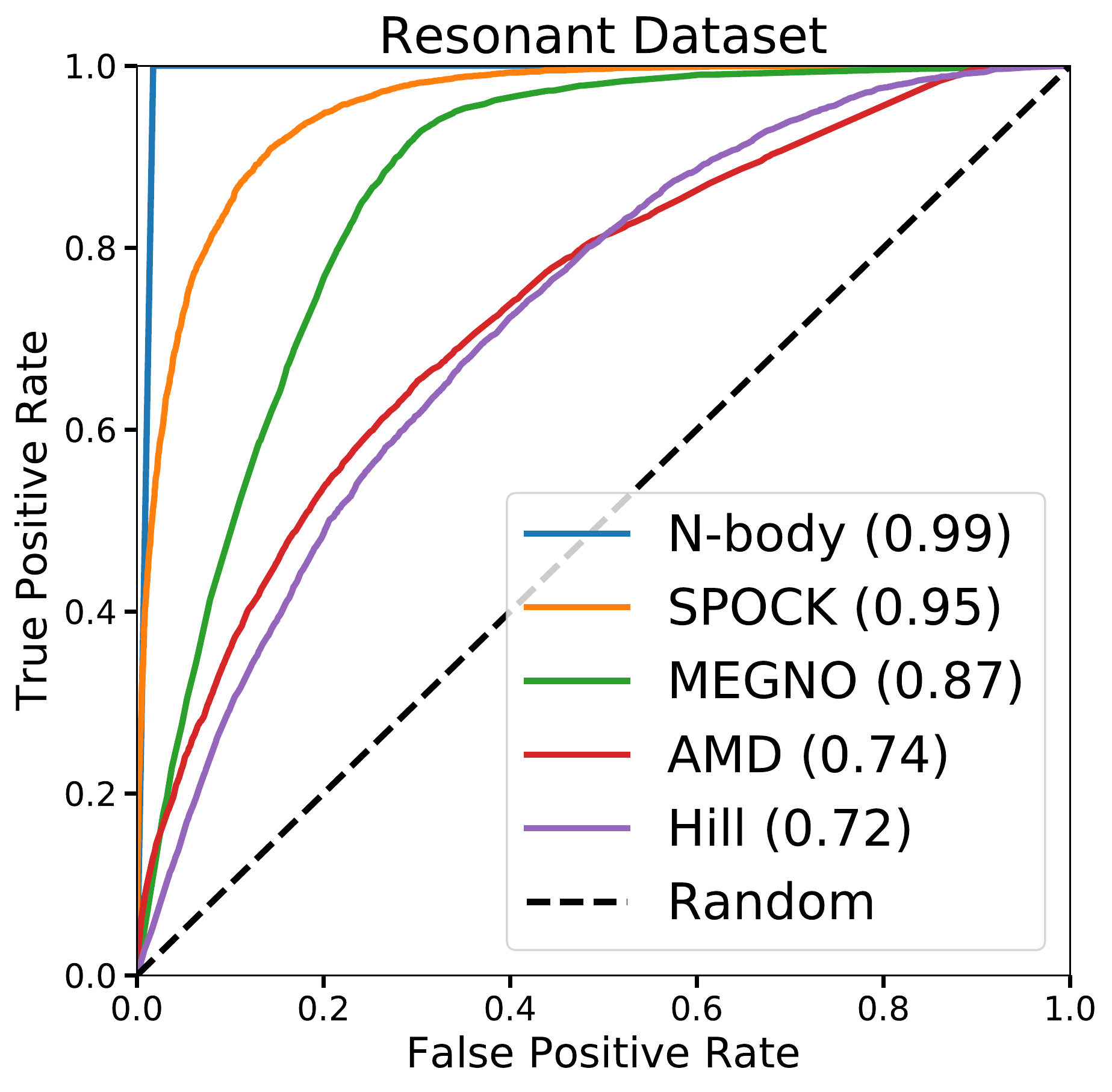}
\caption{Comparison of the performance of SPOCK against previous models on a holdout test set from our training data ({\it Materials and Methods}). Plots True Positive Rate (TPR: fraction of stable systems correctly classified) vs. False Positive Rate (FPR: fraction of unstable systems misclassified as stable). All models can trade off TPR vs. FPR by adjusting their threshold for how conservative to be before labeling a system as stable. The area under the curve (AUC) for each model is listed in the legend in parentheses. A perfect model would have a value of unity, and random guessing (dashed black line) would have AUC=0.5. The blue N-body curve gives an empirical estimate of the best achievable performance on this dataset. At an FPR of 10\%, SPOCK correctly classifies 85\% of stable systems, MEGNO 49\%, AMD 36\%, and Hill 30\%. For a discussion of the various models, see the main text.
\label{summary}}
\end{figure}

Using a holdout test set of $\approx 20,000$ integrations, we present in Fig.\:\ref{summary} the various models' Receiver Operator Characteristic (ROC) curves.
ROC curves plot the classifiers' true-positive rate (TPR, the fraction of stable systems correctly identified) vs. the false-positive rate (FPR, the fraction of unstable systems incorrectly labeled as stable).
Each model returns an estimated probability of stability and can trade off TPR vs. FPR by adjusting its threshold for how conservative it is before it labels a system as stable.
A perfect model would lie in the top left corner, and random guessing would follow the dashed black line. 

Different applications will have different requirements.
For example, stability considerations could be important in inferring the underlying multiplicity distribution of planets in multi-planet systems detected using short time baselines of observation (e.g., with the TESS mission).
In that limit it becomes important to account for the fact that it becomes harder to stably fit higher multiplicity systems into a short observing window.
Such studies estimate the underlying planet occurrences by correcting for various observation biases, typically injecting fake planetary systems into data to model the detection efficiency across the parameter range.

Injecting self-consistent, stable multi-planet configurations requires a low FPR.
If a system is unstable, one wants to be confident that it will be labeled as unstable and thrown out of the analysis.
If one decided that a 10\% false-positive was acceptable, one could read off the corresponding TPR from Fig.\:\ref{summary}.
N-body would correctly label all stable systems, while SPOCK correctly identifies $85\%$. 
MEGNO, AMD, and Hill are not competitive, with TPR values $\leq 50\%$.
MEGNO and SPOCK are roughly a factor of $10^5$ times faster than N-body, while AMD and Hill-sphere-separation models are effectively instantaneous since they are calculated directly from the initial conditions.

It is important to note that this is an unusually demanding test dataset, asking models to make predictions at sharp resonances where the dynamical behavior changes drastically with small changes in parameters (Fig.\:\ref{datasets}). 
In reality, our Solar System and most exoplanet systems are not close to such MMRs \citep{Fabrycky14}, so one should expect the performance on typical systems to be better for all models than what is shown in Fig.\:\ref{results}.
This approach of focusing on the most problematic resonant systems differs from the more uniform phase space coverage used in previous work, and we expect should yield more robust, generalizable models with fewer training examples.
Conversely, the generalization of such a model trained at sharp resonances to the remaining phase space is a strong test of whether MMRs are indeed dominantly responsible for instabilities in compact planetary systems (see Sec.\:\ref{results}\ref{uniform}).

We now consider why previous models performed poorly.
First, while the Hill sphere separations are demonstrably important quantities \citep{Chambers96, Smith09, Obertas17}, they do not carry any information on other important parameters like the orbital eccentricities.
One therefore should not expect a simple two-parameter classifier to yield accurate predictions, particularly near resonances where the behavior depends sensitively on combinations of several different orbital elements.

Second, AMD stability has been shown to be useful in compact {\it two} planet systems \cite{Laskar17, Petit17}, and can be related to the analytical Hill stability limit in such systems \citep{Petit18}.
While it still retains important dynamical information in the 3+ planet case, we see that by itself it is a poor discriminant of stability.
The most obvious problem given our MMR dataset is that AMD stability applies in the secular limit, where the effects of MMRs are ignored.
As \citep{Laskar17, Hadden19} argue, while MMRs alter the AMD, they tend to induce oscillations that average out over a resonant cycle.
However, this is only true for an isolated MMR; once several resonances overlap and motions become chaotic, AMD is not necessarily conserved.
While this is not a concern for two-planet systems in the AMD-stable region \citep{Petit18}, our integrations show empirically that there are many opportunities for MMR overlap in compact systems with three or more planets, and AMD stability is no longer a stringent criterion.

One might argue that this is asking more from AMD stability than it offers, given that it is supposed to be a worst-case scenario estimate.
It only guarantees stability if the total AMD is below the value needed for collisions.
Above the critical AMD, collisions are {\it possible}, but AMD stability makes no prediction one way or another.
However, even if we only consider the $\approx 19\%$ of systems in our resonant test set that AMD guarantees are stable, only $\approx 49\%$ actually are.

Finally, for the MEGNO model, a small fraction ($\approx 2\%$) of the systems it found to be chaotic (taken as a value of MEGNO after $10^4$ orbits > 2.5) are nevertheless stable.
Even if a system is chaotic (i.e., nearby initial conditions diverge exponentially), it still needs enough time for the eccentricities to diffuse to orbit-crossing values.
For example, the GJ876 system has a Lyapunov (chaotic) timescale of only about $7$~years, despite the system being of order a billion years old \citep{Batygin15}.
Determining that an orbit is chaotic is therefore strongly informative, but not sufficient to determine long-term stability. 
More problematically, $55\%$ of the systems with MEGNO values consistent with being regular (non-chaotic) were, in fact, unstable.
This is because MEGNO can only measure chaos on the timescale of the short integration, so systems with Lyapunov times longer than $10^4$ orbits can nevertheless go unstable and be misclassified by MEGNO alone.
In summary, determining that an orbit is chaotic with MEGNO in a short integration is typically a reliable indicator that the system is not long-term stable, but a MEGNO value consistent with a regular orbit is not a robust measure of long-term stability.

By combining MEGNO with features capturing the MMR dynamics, SPOCK substantially improves on these previous models.

\subsection{Generalization to Uniformly Distributed Systems} \label{uniform}

An important concern with machine learning models is whether they will generalize beyond the training dataset.
Since there are no theoretical generalization bounds for modern techniques like decision trees and neural networks, measuring generalization to a holdout set and out-of-distribution data is essential.
In particular, have they learned something meaningful about the underlying physics, or have they simply memorized the particulars of the training set?
We perform two empirical tests.

First, we generate a complementary dataset of 25,000 uniformly sampled configurations, spanning a continuous range in period ratios, and not necessarily close to MMRs.
This is more representative of typical exoplanet systems that have been discovered, with one important difference. 
We choose our sampling ranges to yield roughly a comparable number of stable and unstable configurations ($\approx 40\%$ were stable), while observed systems are naturally biased toward stable regions of phase space since unstable configurations are short-lived.

\begin{figure}
\centering
\includegraphics[width=\linewidth]{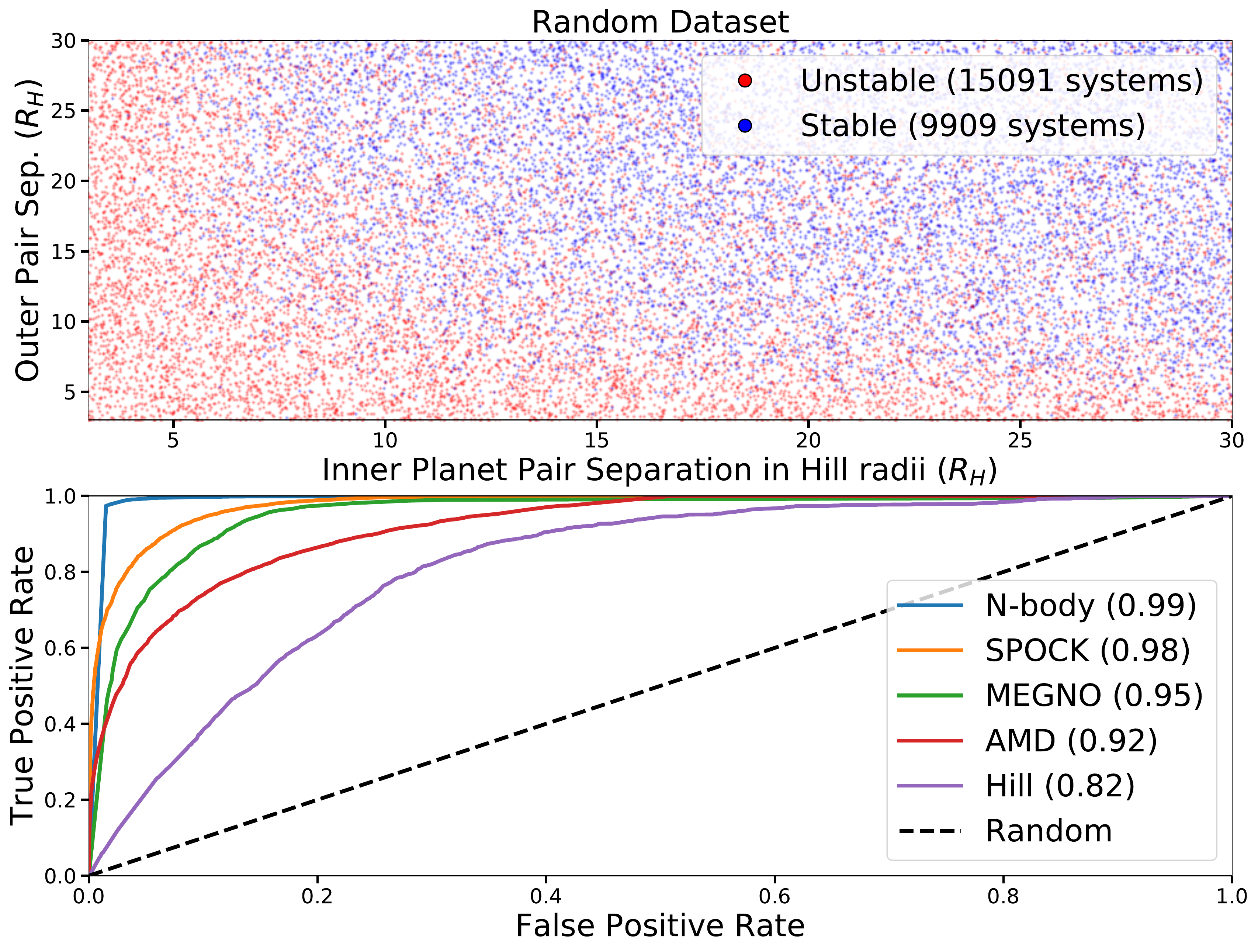}
\caption{Top panel shows the projection of our `random' dataset of 25,000 configurations, uniformly sampling the separation between adjacent planets. 
This is more representative of discovered compact exoplanet systems than the resonant training dataset.
Bottom panel is analogous to Fig.\:\ref{summary}.
At an FPR of $10\%$, SPOCK correctly classifies $94.2\%$ of stable systems, despite having been trained on a different dataset of near-resonant configurations (Fig.\:\ref{datasets}).
}
\label{randomperf}
\end{figure}

The procedure and parameter ranges are the same as in our training set, except we now sample {\it all} planets' orbital elements like we did the third planet above (separations uniform from [0,30] $R_H$, eccentricities log-uniform up to the value at which orbits cross, and all phases uniformly from $[0,2\pi]$).
We plot the projection of this ``random dataset'' into the space spanned by the Hill-radius separations between adjacent planets in the top panel of Fig.\:\ref{randomperf}, where they uniformly fill the plane.

It is easier to predict stability on this dataset for at least two important reasons.
First, most configurations are not particularly close to strong MMRs where the dynamical behavior changes sharply.
Second, while in the resonant training dataset we restricted ourselves to systems that survived longer than our short integrations of $10^4$ orbits, in reality many unstable configurations will be extremely short-lived. 
In our random dataset, we therefore allow for any instability time, which is more representative of typical applications.
This will in particular significantly boost the performance of both the SPOCK and MEGNO models, since they will be able to confidently classify the configurations that go unstable within the span of their short integrations.

We plot the performance of all models (trained on the resonant dataset, Fig.\:\ref{datasets}) on our random dataset in Fig.\:\ref{randomperf}.
All models improve as expected, particularly SPOCK and MEGNO.
At an FPR of $10\%$, N-body correctly classifies $99.8\%$ of systems, SPOCK $94\%$, MEGNO $87\%$, AMD $74\%$ and Hill $39\%$.
Over the range of FPRs in Fig.\:\ref{randomperf}, SPOCK correctly labels approximately half of the systems misclassified by MEGNO.

The fact that our SPOCK classifier, trained on a discrete set of near-resonant systems (Fig.\:\ref{datasets}) performs strongly on this uniform dataset supports our assertion that instabilities in compact multiplanet systems are dominantly driven by MMRs. 
If instead we let SPOCK train on $80\%$ of our random dataset and test on the remaining $20\%$, the TPR quoted above only rises by $\approx 2\%$, suggesting our model can robustly classify a wide range of compact three-planet systems.

\subsection{Generalization To Higher Multiplicity Systems} \label{5p}

Influenced by previous work \citep{Chambers96}, we hypothesized that the simplest case for understanding instabilities within $10^9$ orbits in multi-planet systems is that of three planets.
A natural question is therefore how well our model, trained on three-planet systems, generalizes to higher-multiplicity cases. 

We test our model's generalization on previously published numerical integrations of five equal-mass planets on co-planar, equally spaced, and initially circular orbits \citep{Obertas17}.
This is in stark contrast to our systems of three unequal-mass planets on mutually inclined, unevenly spaced, and eccentric orbits.
Indeed, the integrations only varied the separation between adjacent planets, corresponding to a diagonal line from the bottom left to the top right of Fig.\:\ref{datasets}). 
This passes through the many intersections between vertical and horizontal MMR lines, where our analytic transformations (assuming the influence of only a single MMR) are most problematic.
All other parameters were fixed, so there are very few examples in our much higher dimensional training set that fall near this 1-dimensional line, rendering it a particularly stringent test of our model.
In particular, if SPOCK had simply memorized the particulars of our training set, it should not be able to effectively predict on systems drawn from a very different configuration distribution.

As a simple prescription for predicting stability on 4+ planet systems using SPOCK, we feed all adjacent trios to our three-planet classifier and retain the lowest stability probability.
This is a simplification, but should provide a reasonable approximation in cases where instabilities are driven by perturbations on an MMR between a particular pair of planets. We argue this typically is the case in compact systems.

\begin{figure}
\centering
\includegraphics[width=\linewidth]{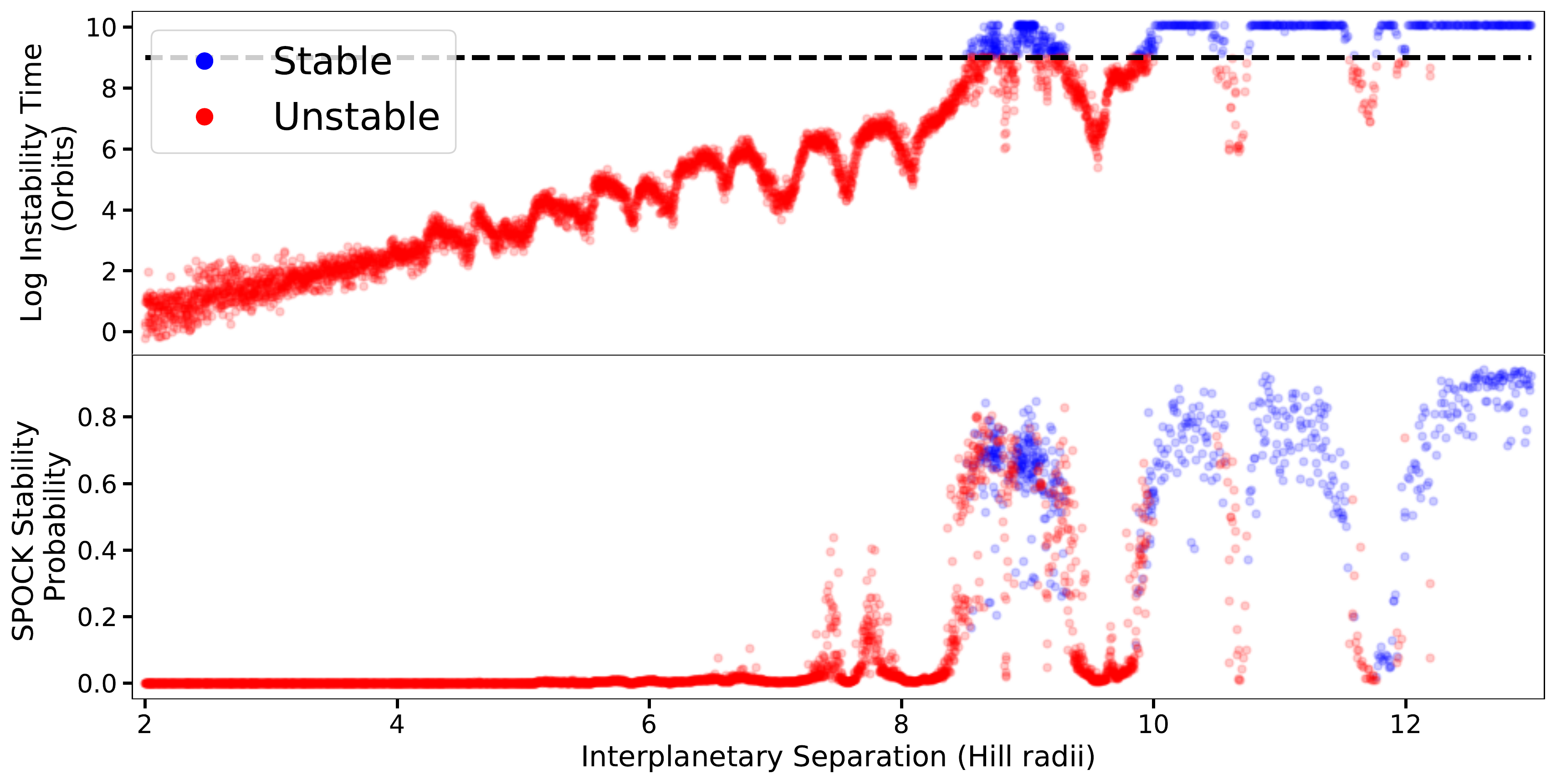}
\caption{Generalization of SPOCK (trained on eccentric, inclined, unequally spaced, unequal-mass three-planet systems in and near MMRs) to integrations from \cite{Obertas17} of initially circular, co-planar, equally spaced, equal-mass five-planet systems.
The separation between all adjacent pairs of planets increases along the x-axis.
Top panel shows the instability times measured by direct integration, with dips corresponding to MMRs \citep{Obertas17}.
Bottom panel shows the stability probability predicted by SPOCK.
Taking the probability of stability threshold of 0.34 used in Sec.\:\ref{results}\ref{holdout}, the true positive rate is 94\% and the false positive rate 6\% on this dataset.
}
\label{alysa}
\end{figure}

In Fig.\:\ref{alysa}, the top panel shows the instability time recorded by the 17500 N-body integrations of \cite{Obertas17}, plotted against the separation between adjacent planets (normalized by their mutual Hill radius (Eq.\:\ref{RH})).
As above, we color-code systems that went unstable within $10^9$ orbits as red, and stable systems as blue.
While above we considered binary classification as stable or unstable, in the bottom panel we now plot the {\it probability} of stability estimated by SPOCK for each of the initial conditions in order to better visualize the model's sensitivity to the structure in instability times visible in the top panel (each point along any of the SPOCK ROC curves above corresponds to the TPR and FPR obtained when setting a particular stability probability threshold).

We see that SPOCK gives reliable results, despite having been trained on very different configurations of resonant and near-resonant configurations of fewer planets.
Figure \ref{alysa} also shows that SPOCK recognizes each of the dips in instability times in the top panel, which correspond to the locations of MMRs \citep{Obertas17}, and adjusts its stability probability accordingly.
Misclassifications are largely limited to the boundaries between stable and unstable configurations in the top panel.
We note that near this boundary classification is ambiguous---some of these systems would also be ``misclassified'' by direct N-body integrations.
Using the same threshold as in Sec.\:\ref{results}\ref{holdout} (chosen to yield an FPR on our resonant holdout set of 10\%), the TPR across this test set of 17500 integrations is 94\%, with an FPR of 6\%.
The fact that our model trained on three-planet systems generalizes to higher multiplicities supports our assertion at the outset that planet trios are prototypical cases that can be extended to higher numbers of planets.

\subsection{An Application} \label{application}

\begin{figure}
\centering
\includegraphics[width=\linewidth]{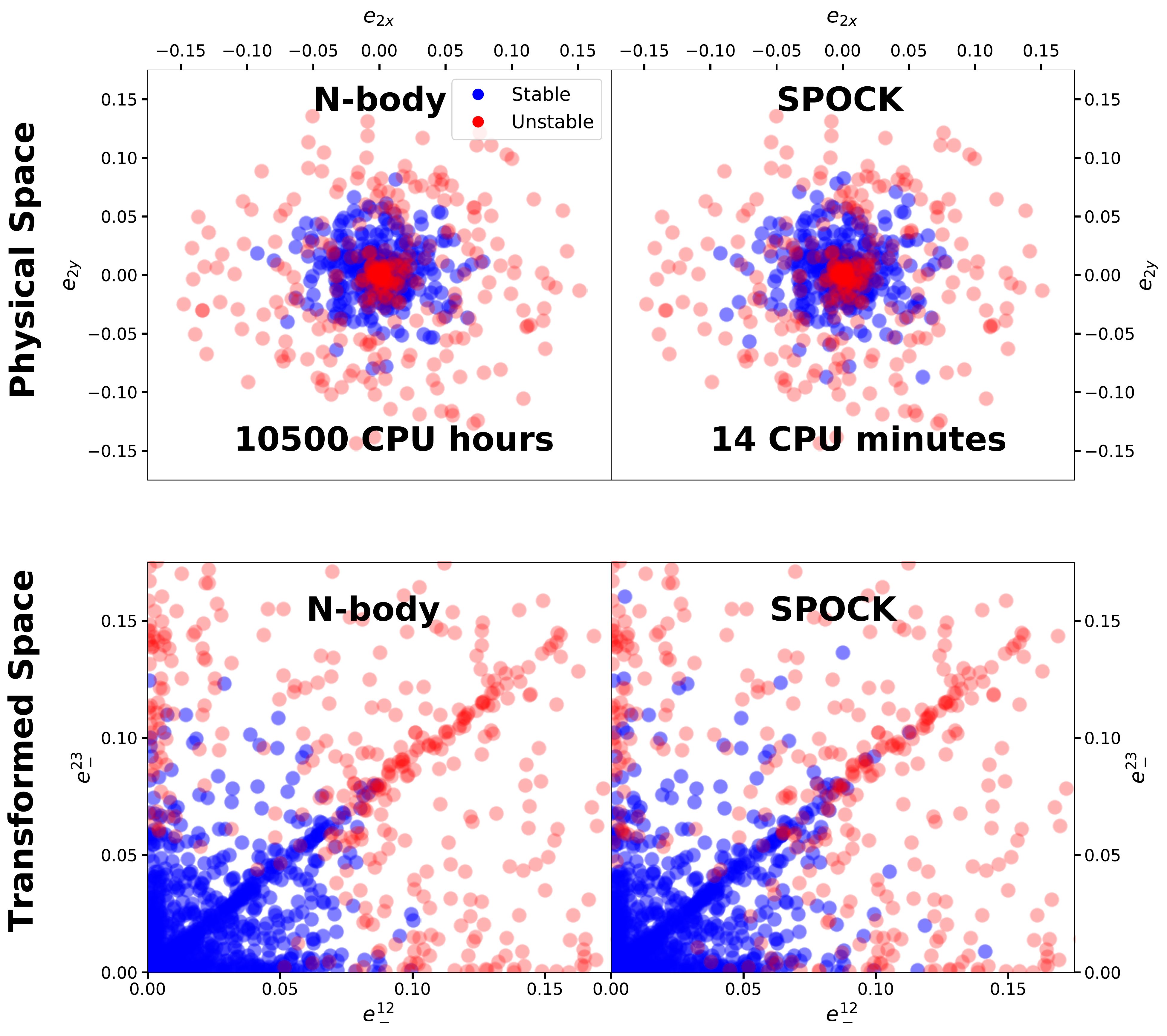}
\caption{Top panels are polar plots of the middle planet's eccentricity vector in the Kepler 431 compact three-planet system (distance from origin is the eccentricity; polar angle the direction toward pericenter).
Top left panel color codes 1500 stable and unstable configurations through direct integration, while top right panel shows the stability predictions from SPOCK---96\% of points agree.
The bottom panel projects the configurations into the transformed space used by our model features (see text).
The stability boundary separates much more cleanly in this space that incorporates our understanding of the resonant dynamics, showing visually why our engineered features help the algorithm's performance.
\label{kep431}}
\end{figure}

As an example, we considered the characterization of four observed compact three-planet systems (Kepler-431, Kepler-446, EPIC-2108975, and LP-358-499), none of which are near strong MMRs.
We again focus on compact systems, since we should be able to reject a larger range of masses and orbital eccentricities for these more delicate configurations.
All four systems gave similar results, so we focus on Kepler-431, a system of three transiting Earth-size planets (which gave the second-worst performance).

The planetary transits across the host star strongly constrain the planets' orbital periods and physical sizes. The masses and especially the orbital eccentricities remain highly uncertain.
As a simple exercise, we sample the planetary masses from a mass-radius relationship \citep{ChenKip16} and sample eccentricities log-uniformly between [$10^{-4},0.18$] for each of the three planets independently (with the upper limit representing the value at which the inner two orbits would cross). 
Since these are transiting planets, we draw inclinations from an edge-on configuration uniformly from $10^{-3}$ radians to the angular size of the star as seen from the planet, $R_\star/a_i$ (with $R_\star$ the stellar radius and $a_i$ the ith planet's semimajor axis).
All remaining angles are drawn uniformly from $[0,2\pi]$, and we assume a stellar mass of $1.07$ solar masses.
We draw 1500 configurations in this way, and for each one run both direct N-body integrations and our SPOCK classifier.

Adopting the same stability probability threshold from Sec.\:\ref{results}\ref{holdout}, we obtain the results plotted in Fig.\:\ref{kep431}.
To visualize the phase space, in the top row we provide polar plots of the middle planet's eccentricity vector (with the distance from the origin giving the eccentricity, and the polar angle the direction toward pericenter).
The top left panel color codes stable and unstable configurations obtained through direct N-body.
The top right panel shows the predictions from SPOCK, yielding an FPR of 9\% and TPR of 97\%.

While the expected trend of instability toward high eccentricities is born out in the top panel, many unstable configurations remain near the origin at zero eccentricity due to other system parameters not visible in this projection.
However, by developing a classifier with a comparatively small number of physically motivated features, we can gain insight into the stability constraints by projecting the configurations onto the transformed resonant space used by the model. 
In the bottom panel we consider the eccentricity modes $e_-$ (Eq.\:\ref{eminus}) that dominate the MMR dynamics between each adjacent pair of planets (first and second planet on the x-axis; second and third on the y-axis). 
We see that our feature space incorporating our analytical understanding of the resonant dynamics much more cleanly separates the stable and unstable systems, even in this 2-D projection.
This both visually shows how our engineered features help the algorithm's performance, and clarifies the particular combinations of parameters specifically constrained by stability.

Finally, we note that in this closely packed system, stability is indeed constraining.
We constrain the free eccentricities\footnote{We quote free eccentricities typically quoted for TTV constraints $Z \approx |{\bf e_-}|/\sqrt{2}$ (see {\it Materials and Methods}) for a more direct comparison} of the inner and outer pair of planets to be below $0.051$ and $0.053$, respectively (84th percentile limit).
Such eccentricity limits, which constrain the degree of dynamical excitation in the system's past \citep{Chatterjee08, Juric08}, are significantly stronger than those inferred from radial velocity \citep[e.g.,][]{Butler06} or transit duration measurements \citep{vanEylen15, Xie16} for such low-mass planets, which dominate the population \citep[e.g.,][]{Burke15}.
Within a factor of a few, this approaches the exquisite constraints achievable by modeling transit timing variations (TTVs), which are typically only measurable when planets are close to strong MMRs,
with accurate photometry, and with long observation baselines \citep{Holman05}.
In particular, TTVs are not detected in any of the four Kepler systems we considered.
TTV modeling has been an extremely productive method with the long observation baselines of the Kepler mission \citep[e.g.,][]{Jontof14, Hadden17}.
However, the much shorter observing windows of Kepler's successor, the Transiting Exoplanet Survey Satellite (TESS), implies that only $\sim 10$ planets are expected to be constrained by TTVs \citep{Hadden2019_tess} during its prime mission.
This places stability constrained characterization as a powerful complementary method for understanding multi-planet systems.

\subsection{Limits} \label{limits}
Finally, we present an instructive case where SPOCK fails, for systems constrained by above-mentioned TTVs. 
Transiting planets that do not interact with one another would pass in front of their host stars like perfect clocks with a constant orbital period.
However, their mutual gravitational tugs can cause transit times to periodically pull ahead and fall behind.
This is a particularly strong effect near MMRs, which induce sinusoidal TTVs\citep{Holman05, Lithwick12}.

We considered six systems that exhibit TTVs, and in particular, the three-planet Kepler-307 system \citep{Jontof16} (outermost planet only a candidate).
In all cases, the transit times have been fit to infer planet masses and orbital parameters with Markov Chain Monte Carlo (MCMC).
We choose to sample 1500 configurations from the resulting posterior, and again run N-body integrations to compare with SPOCK predictions as in Sec.\:\ref{application}.

\begin{figure}
\centering
\includegraphics[width=\linewidth]{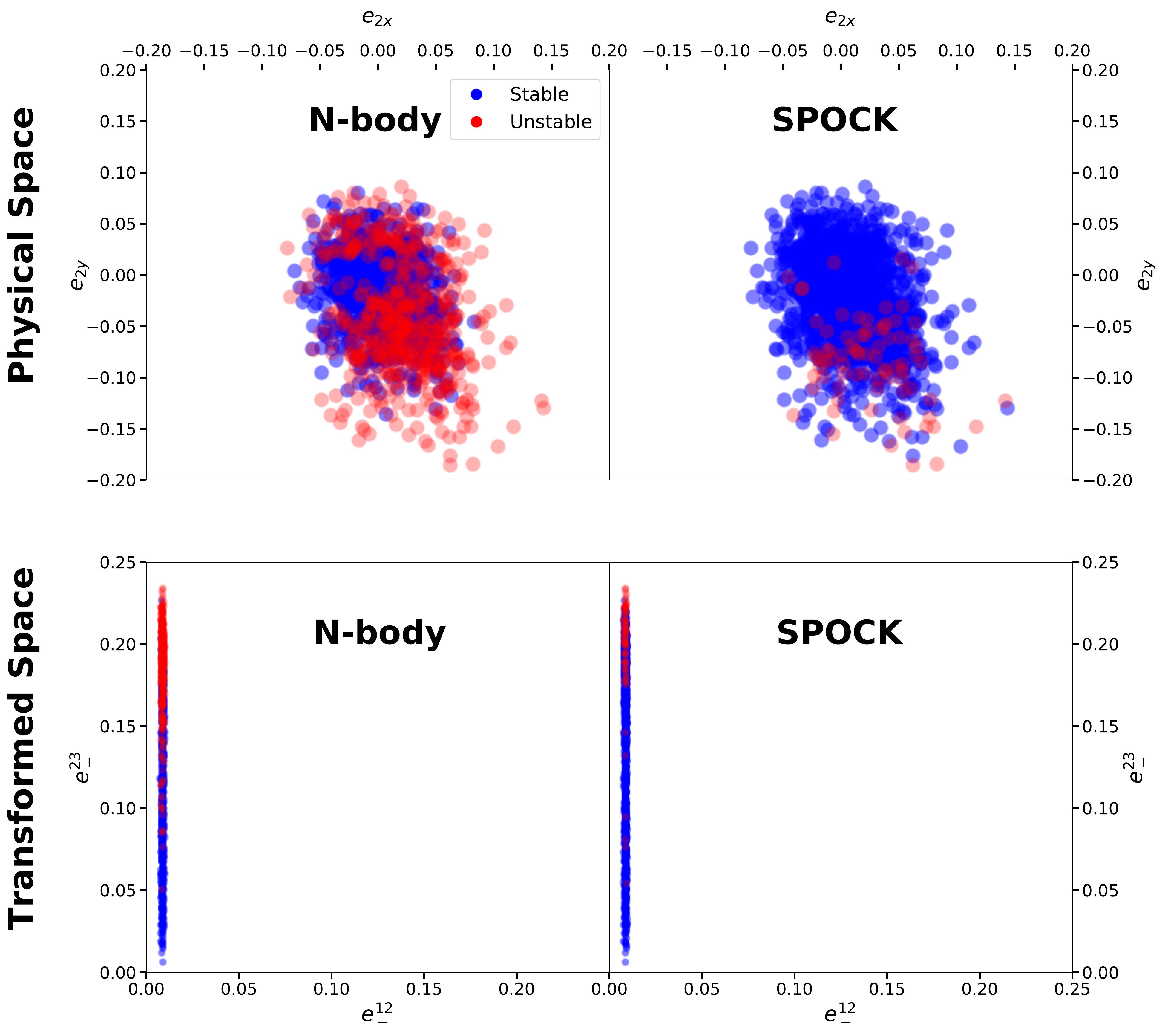}
\caption{
Case of the 3-planet Kepler-307 system, where SPOCK predictions fail. We sample configurations (points) from the posterior of a Markov Chain Monte Carlo fit to the observed transit timing variations (TTVs) in the system.
Top and bottom rows are analogous to Fig.\:\ref{kep431}. 
Transforming to the 2-body resonant variables used by SPOCK in the bottom row shows the reason for the poor performance. 
The previous TTV fit has constrained the eccentricity mode dominantly driving the MMR dynamics of the inner two planets to an extremely narrow range.
This leaves points only along a direction that does not strongly influence the short integrations we use to generate features.
\label{KOI1576}
}
\end{figure}

Interestingly, SPOCK fails on all of them.
In the case of Kepler-307, the FPR is 87\% (Fig.\:\ref{KOI1576}).
An important cost to consider with complex models is the difficulty in diagnosing problems such as these when they come up.
Our original SPOCK model generated 60 summary features from short integrations, and in fact slightly outperformed our final adopted model on the holdout set in Fig.\:\ref{summary}.
However, we chose to trade these marginal performance gains for the improved interpretability of our smaller set of 10 physically relevant features, and this reveals the reason for the poor performance in Fig.\:\ref{KOI1576}.

The inner two planets in this system are near a 5:4 MMR (period ratio $\approx 1.255$), while the third planet is significantly further separated (period ratio between the outer two planets $\approx 1.79$).
As mentioned above, the MMR dynamics between a pair of planets are driven by a particular combination of the orbital eccentricities $e_-$ (Eq.\:\ref{eminus}).
In this case, because the observed TTVs are driven by a 5:4 MMR between the inner two planets, the TTVs observed in the data specifically constrain this planet pair's $e_-$ mode.
If we again transform the space in the top row of Fig.\:\ref{KOI1576} to that spanned by the $e_-$ modes for both adjacent pairs like in Fig.\:\ref{kep431}, we see that the sample of configurations collapses to a thin vertical line.

The problem is therefore that while SPOCK would typically help to constrain $e_-$ by ruling out unstable values, the MMR-driven TTVs have {\it already} allowed the MCMC fit to narrow down the $e_-$ mode for the inner pair of planets to an exquisitely narrow range of $0.0088 \pm 0.0004$.
Thus, samples from the MCMC posterior have already removed configurations along directions in which SPOCK has strong discerning power, leaving only points along directions that are difficult to separate from the short integrations.
The `MEGNO' model similarly fails with an FPR of $57\%$.

If we constrained the above system blindly, across the full range of possible eccentricities, SPOCK's performance would be comparable to the results in Sec.\:\ref{results}\ref{holdout}.
In this case, however, observational data has strongly constrained the planets' resonant dynamics, leaving only configurations with very similar SPOCK features, and leading to unreliable predictions.
Presumably, improved models would incorporate additional features that better separate the stable and unstable configurations in Fig.\:\ref{KOI1576}.
Such SPOCK failures should be rare, but TTV-constrained configurations are an important, instructive counterexample.
One can test for such situations empirically by looking for clustering of configurations in SPOCK's feature space.
An important advantage of SPOCK's physically meaningful features is that it facilitates the interpretation of any such clusterings.

\section{Conclusion}

We have presented the Stability of Planetary Orbital Configurations Klassifier (SPOCK), a machine learning model capable of classifying stability of compact 3+ planet systems over $10^9$ orbits.
SPOCK is up to $10^5$ times faster than direct N-body integration, and is significantly more accurate (Figs.\:\ref{summary} and \ref{randomperf}) than stability predictions using AMD stability \citep{Laskar17}, Hill-sphere separations \citep[e.g.,][]{Chambers96, Zhou07, Quillen11}, or the MEGNO chaos indicator \citep[e.g.,][]{Migaszewski12}

This computationally opens up the stability constrained characterization of compact multi-planet systems, by rejecting unphysical, short-lived candidate orbital configurations.
In the Kepler 431 system with three tightly packed, approximately Earth-sized planets,we constrained the free eccentricities of the inner and outer pair of planets to both be below $0.05$ (84th percentile upper limits). 
Such limits are significantly stronger than can currently be achieved for small planets through either radial velocity or transit duration measurements, and within a factor of a few from transit timing variations (TTVs).
Given that the TESS mission's typical 30-day observing windows will provide few strong TTV constraints \citep{Hadden2019_tess}, SPOCK computationally enables stability constrained characterization as a productive complementary method for extracting precise orbital parameters in compact multi-planet systems.

Our training methodology and tests also clarify the dynamics driving instabilities in compact exoplanet systems.
Our model, trained solely with configurations in and near MMRs, accurately predicts instabilities within $10^9$ orbits across the full phase space of typical compact systems (Sec.\:\ref{results}\ref{uniform}).
This is strong confirmation that {\it rapid} instabilities, on timescales much shorter than observed systems typical $\sim 10^{11}$-orbit dynamical ages, are dominantly driven by the overlap of MMRs \citep{Wisdom80, Quillen11, Obertas17}.

Instabilities can also occur on longer timescales through the overlap of secular resonances.
As opposed to MMRs between planets' orbital rates, secular resonances represent commensurabilities between the much slower rates at which orbits precess. 
This is the case for our solar system, which has a dynamical lifetime $>10^{10}$ orbits \citep{Laskar09,Lithwick11secularchaos, Batygin15}.
SPOCK is not trained to detect such slow instabilities, but self-consistently classifies the solar system as stable over $10^9$ orbits.

Recent work \citep{Volk20} suggests that instabilities in compact systems are driven through the overlap of such secular resonances.
While this may seem in tension with our focus on MMRs, this paints a self-consistent picture.
Short-lived configurations eliminate themselves, rearranging and dynamically carving out the distribution of planetary systems that survive to the present day.
This idea has been advanced from several perspectives \citep{Barnes04, Volk15, Pu15, Tremaine15, Izidoro17}.
MMR-driven instabilities happen quickly compared to the typical ages of observed systems, leaving today only systems that destabilize through slower secular instabilities.
This also clarifies that secular analyses such as AMD stability are valuable dynamical classifications for observed systems \citep{Laskar17}, despite their poor identification of short-term instabilities (Sec.\:\ref{results}).

We also showed that short-term, MMR driven instabilities are local, as expected from the lack of strong MMRs beyond period ratios of 2:1 \citep{Murray97, Quillen11}.
In particular, we showed that our model, trained on three-planet systems, can be applied to adjacent trios of planets in higher multiplicity systems to classify stability over $10^9$ orbits.
This implies that stability constrained characterization is robust against distant, unseen planets (an important consideration for detection methods heavily biased against finding such bodies).
This is not the case for longer timescale secular instabilities.
For example, exodynamicists detecting only the inner solar system would infer a much more stable system than is actually the case \citep[e.g.,][]{Lithwick11secularchaos, Batygin15}.

By identifying the dominant dynamics driving the instabilities we aimed to classify, and incorporating this directly into both the training set and the set of features used for our machine learning model, we have trained a robust classifier of multi-planet stability over $10^9$ orbits.
This approach also allowed us to both test our assumptions, and understand regions of phase space where SPOCK should fail.
This can be a useful blueprint for exploiting the often extensive domain knowledge available in scientific applications of machine learning.

We make our $\approx 1.5$-million CPU-hour training set publicly available (\url{https://zenodo.org/record/3723292}), and provide an open-source package, documentation and examples (\url{https://github.com/dtamayo/spock}) for efficient classification of planetary configurations that will live long and prosper.

\matmethods{ \label{methods}

\subsection{Resonant Dataset} \label{resdataset}

We initialize our near-resonant pair of planets by first identifying all the first ($n:n-1$) and second ($n:n-2$) order MMRs in the range from [3.5, 30] mutual Hill radii ([3.5, 60] mutual Hill radii for non-adjacent planets)\footnote{the lower limit is the Hill stability limit \citep{Gladman93} was chosen to avoid immediate instabilities.}, which represent the strongest set of resonances \citep{Murray99}.
We then randomly assign the pair to one of the resonances in this list.

A pair planets near a first-order $n:n-1$ MMR, whose orbits evolve in a twelve-dimensional phase space, can to excellent approximation be reduced to a two-dimensional dynamical system through a transformation of variables involving several conserved quantities \cite{Sessin84}.
This transformation has recently been generalized for $n:n-2$ and higher-order MMRs \citep{Hadden19} \citep[see also][]{Hadden18}.

First, at low eccentricities, the eccentricity and inclination evolution decouple. 
Therefore, we sample the planets' orbital inclinations and orbital plane orientations randomly as described in Sec.\:\ref{trainingset}.
For the eccentricities, the intuition is that only one mode (or combination) matters. 
In particular, a combination $\bf{Z_+}$, approximately the ``center-of-mass'' eccentricity vector, is conserved, while $\bf{Z}$ drives the dynamics \citep[e.g.,][]{Hadden19},
\begin{equation} \label{modes}
\sqrt{2}{\bf Z} \approx {\bf e}_2 - {\bf e}_1 \equiv {\bf e}_-, \:\:\:\:\: {\bf Z_+} \approx \frac{m_1{\bf e}_1 + m_2{\bf e}_2}{m_1+m_2} \equiv {\bf e}_+ \approx const.,
\end{equation}
where ${\bf e}_1$ and ${\bf e}_2$ are the planets' orbital eccentricity vectors, and $m_1$ and $m_2$ are their respective masses.
The {\bf $Z$} and {\bf $Z_+$} vectors incorporate additional coefficients on the eccentricity vectors that depend on the particular MMR.
With the exception of the 2:1 MMR, these coefficients are within $\approx 10$\% of unity \citep{Deck13}, and converge to one as the period ratio shrinks \citep{Hadden19}.
Models trained with ($Z$, $Z_+$) exhibit similar performance to ones trained with ($e_-$, $e_+$), so we adopt the latter to avoid discontinuities in our features near discrete resonances.

Finally, one can combine the period ratio and other orbital parameters to define a metric for proximity to the resonance.
The closer the system is to the resonance, the higher ${\bf e}_-$ will be forced by the MMR.
We sample the relevant range in proximities to resonance by drawing the eccentricity forced by the resonance $e_{\rm forced}$ (see Fig.\:\ref{phaseportrait}) log-uniformly from the minimum value (predicted by the analytic theory to yield an isolated resonant island \citep{Hadden19}) to the nominal orbit-crossing value $e_{\rm cross}$ (Eq.\:\ref{ecross}).
Choosing $e_{\rm forced}$ defines a family of different possible trajectories, depending on the remaining initial conditions, which we plot in Fig.\:\ref{phaseportrait}.
The variable on the x-axis is the resonant angle $\phi$, which is the particular combination of the pericenter orientations and phases corresponding to the given MMR---physically, it tracks the position at which conjunctions occur, where planets most strongly interact.
The choice of $e_{\rm forced}$, or proximity to resonance, moves the entire resonant island (bounded by the black curve) up and down.

\begin{figure}
\centering
\includegraphics[width=.99\linewidth]{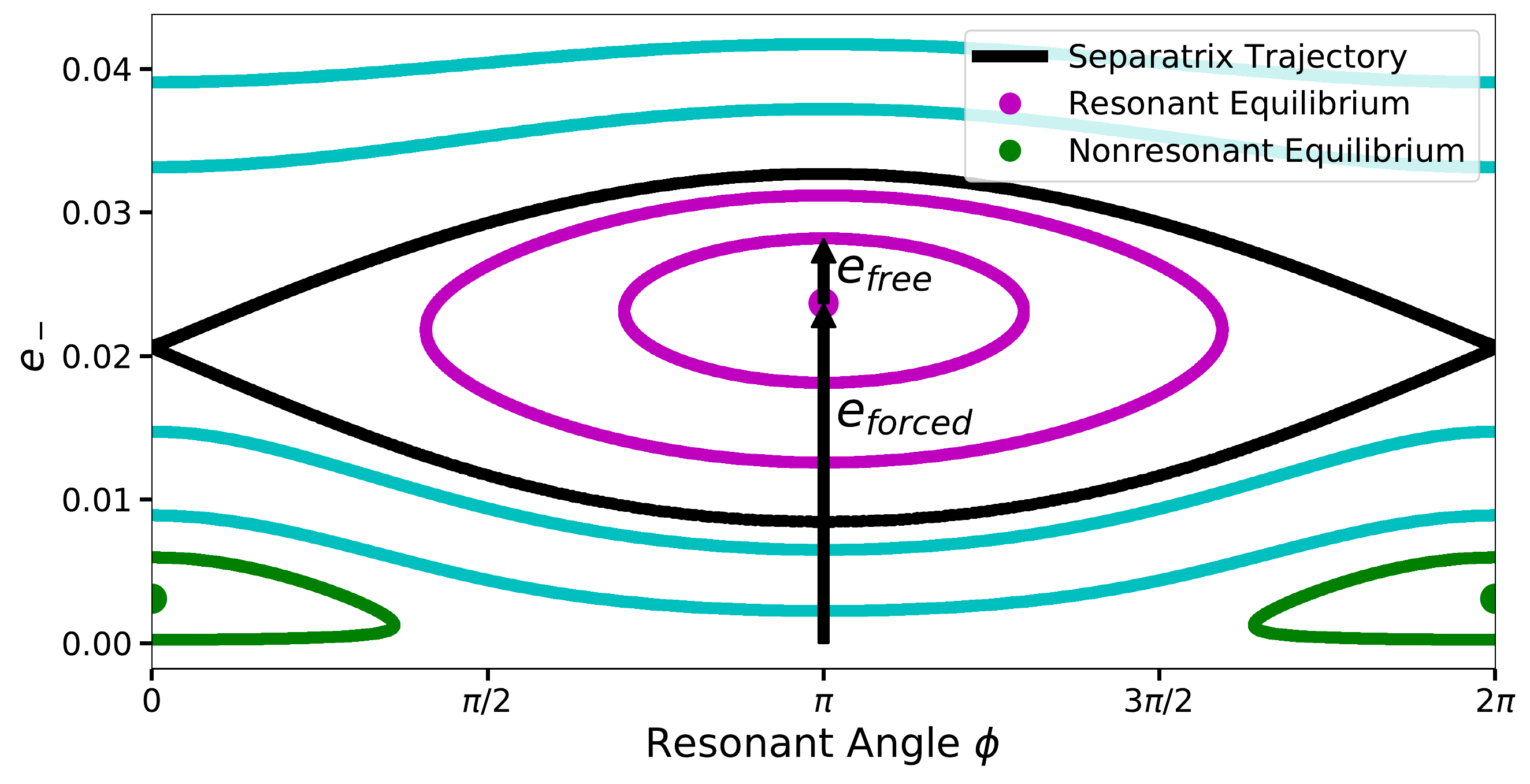}
\caption{Phase portrait of the dynamics near a MMR. 
See {\it Materials and Methods} for discussion.
\label{phaseportrait}
}
\end{figure}

The final important parameter is then the free eccentricity $e_{\rm free}$, which measures the distance from the resonant equilibrium (red dot in the bottom panel of Fig.\:\ref{phaseportrait})
A system initialized with $e_{\rm free}=0$ at the center of the island will remain at the equilibrium.
Configurations initialized with small non-zero values of $e_{\rm free}$ will oscillate (librate) around the equilibrium like a pendulum.
There is a critical value of $e_{\rm free}$ beyond which the system will no longer librate around the equilibrium but instead circulates. 
The boundary trajectory, which forms a `cat's eye' splitting these two types of behaviors, is called the separatrix (black curve). 

To fill in this range of behaviors, we sample $e_{\rm free}$ log-uniformly from $[3\times 10^{-3},3]$ of the distance to the separatrix along $\phi=\pi$.
In this way, the resonant pair of planets spans the range from being in resonance to being outside the resonant region, but still having their dynamics strongly influenced by the MMR.
Many of the planets discovered by the Kepler mission that exhibit transit timing variations lie in the region near (but outside) strong MMRs\citep[e.g.,][]{Hadden16}.
When the drawn $e_{\rm free}$ is large enough to move beyond the bottom of the plot, we wrap around and initialize the system at 
$\phi=0$.
This allows us also to sample the other island shown in green in the bottom panel of Fig.\:\ref{phaseportrait}, which has an equilibrium at small values of $e_-$.
Once the initial conditions are allowed to evolve, they fill the space in Fig.\:\ref{phaseportrait}, as shown by the few plotted sample trajectories.
For the conserved quantity $e_+$, we sample it log-uniformly within the same range as $e_{\rm forced}$. We draw the remaining nuisance angles uniformly from $[0,2\pi]$.

We initialized the resonant pair of planets using the open-source {\tt celmech} package (\url{https://github.com/shadden/celmech}), which is based on\cite{Hadden19}.
{\tt celmech} includes an API for initializing resonant orbital configurations from the above parameters, and we include the scripts and random seeds used to generate our training sets in the repository accompanying this paper.

\subsection{Numerical Integrations} \label{integrations}

All integrations were performed with {\whfast} \citep{ReinTamayo15}, part of the {\reb} N-body package \citep{Rein12}.
We adopted a timestep of $\approx 3.4\%$ of the innermost planet's orbital period.
If any planets' Hill spheres overlapped, the simulation was ended and the instability time recorded.
The specific halting condition is not important \citep{Gladman93}, as once Hill spheres start crossing, the system becomes an orbit-crossing tangle on orbital timescales\footnote{It might still take a long time for small planets close to their host star to find one another and collide \citep{Rice18}; however, in the context of applying stability constraints, we are usually interested in the time to instability defined such that the system architecture becomes inconsistent with the typically observed multi-planet system of approximately planar, concentric near-circular orbits.}.

The integrations analyzed in this work were saved in the \reb {\:\tt SimulationArchive} format, which enables exact, machine-independent reproducibility of results \citep{ReinTamayo17}.
We provide instructions and the scripts necessary to reproduce the figures in the accompanying repository.
}
\showmatmethods{} 

\acknow{}

\showacknow{} 
We are grateful to Antoine Petit and the anonymous reviewers for insightful suggestions that significantly improved the manuscript.
Support for DT was provided by NASA through the NASA Hubble Fellowship grant HST-HF2-51423.001-A awarded  by  the  Space  Telescope  Science  Institute,  which  is  operated  by  the  Association  of  Universities  for  Research  in  Astronomy,  Inc.,  for  NASA,  under  contract  NAS5-26555.
HR has been supported by the NSERC Discovery Grant RGPIN-2014-04553 and the Centre for Planetary Sciences at the University of Toronto Scarborough.

\bibliography{Bib}

\begin{thebibliography}{82}
\providecommand{\natexlab}[1]{#1}
\providecommand{\url}[1]{\texttt{#1}}
\expandafter\ifx\csname urlstyle\endcsname\relax
  \providecommand{\doi}[1]{doi: #1}\else
  \providecommand{\doi}{doi: \begingroup \urlstyle{rm}\Url}\fi

\bibitem[Kolmogorov(1954)]{Kolmogorov54}
Andrey~Nikolaevich Kolmogorov.
\newblock On conservation of conditionally periodic motions for a small change
  in hamilton's function.
\newblock In \emph{Dokl. Akad. Nauk SSSR}, volume~98, pages 527--530, 1954.

\bibitem[M{\"o}ser(1962)]{Moser62}
J~M{\"o}ser.
\newblock On invariant curves of area-preserving mappings of an annulus.
\newblock \emph{Nachr. Akad. Wiss. G{\"o}ttingen, II}, pages 1--20, 1962.

\bibitem[Arnol'd(1963)]{Arnold63}
VI~Arnol'd.
\newblock Proof of a theorem of an kolmogorov on the invariance of
  quasi-periodic motions under small perturbations of the hamiltonian.
\newblock \emph{Russian Mathematical Surveys}, 18\penalty0 (5):\penalty0 9,
  1963.

\bibitem[H{\'e}non(1966)]{Henon66}
M~H{\'e}non.
\newblock Explorationes num{\'e}rique du probleme restreint iv: Masses egales,
  orbites non periodique.
\newblock \emph{Bullettin Astronomique}, 3\penalty0 (1):\penalty0 49--66, 1966.

\bibitem[Celletti and Chierchia(2005)]{Celletti05}
Alessandra Celletti and Luigi Chierchia.
\newblock Kam stability for a three-body problem of the solar system.
\newblock \emph{Zeitschrift f{\"u}r angewandte Mathematik und Physik ZAMP},
  57\penalty0 (1):\penalty0 33--41, 2005.

\bibitem[Mills et~al.(2016)Mills, Fabrycky, Migaszewski, Ford, Petigura, and
  Isaacson]{Mills16}
Sean~M Mills, Daniel~C Fabrycky, Cezary Migaszewski, Eric~B Ford, Erik
  Petigura, and Howard Isaacson.
\newblock A resonant chain of four transiting, sub-neptune planets.
\newblock \emph{Nature}, 533\penalty0 (7604):\penalty0 509--512, 2016.

\bibitem[Gillon et~al.(2017)Gillon, Triaud, Demory, Jehin, Agol, Deck, Lederer,
  De~Wit, Burdanov, Ingalls, et~al.]{Gillon17}
Micha{\"e}l Gillon, Amaury~HMJ Triaud, Brice-Olivier Demory, Emmanu{\"e}l
  Jehin, Eric Agol, Katherine~M Deck, Susan~M Lederer, Julien De~Wit, Artem
  Burdanov, James~G Ingalls, et~al.
\newblock Seven temperate terrestrial planets around the nearby ultracool dwarf
  star trappist-1.
\newblock \emph{Nature}, 542\penalty0 (7642):\penalty0 456--460, 2017.

\bibitem[Steffen et~al.(2013)Steffen, Fabrycky, Agol, Ford, Morehead, Cochran,
  Lissauer, Adams, Borucki, Bryson, et~al.]{Steffen13}
Jason~H Steffen, Daniel~C Fabrycky, Eric Agol, Eric~B Ford, Robert~C Morehead,
  William~D Cochran, Jack~J Lissauer, Elisabeth~R Adams, William~J Borucki,
  Steve Bryson, et~al.
\newblock Transit timing observations from kepler--vii. confirmation of 27
  planets in 13 multiplanet systems via transit timing variations and orbital
  stability.
\newblock \emph{Monthly Notices of the Royal Astronomical Society},
  428\penalty0 (2):\penalty0 1077--1087, 2013.

\bibitem[Tamayo et~al.(2015)Tamayo, Triaud, Menou, and Rein]{Tamayo15}
Daniel Tamayo, Amaury~HMJ Triaud, Kristen Menou, and Hanno Rein.
\newblock Dynamical stability of imaged planetary systems in formation:
  Application to hl tau.
\newblock \emph{The Astrophysical Journal}, 805\penalty0 (2):\penalty0 100,
  2015.

\bibitem[Wang et~al.(2018)Wang, Graham, Dawson, Fabrycky, De~Rosa, Pueyo,
  Konopacky, Macintosh, Marois, Chiang, et~al.]{Wang18}
Jason~J Wang, James~R Graham, Rebekah Dawson, Daniel Fabrycky, Robert~J
  De~Rosa, Laurent Pueyo, Quinn Konopacky, Bruce Macintosh, Christian Marois,
  Eugene Chiang, et~al.
\newblock Dynamical constraints on the hr 8799 planets with gpi.
\newblock \emph{The Astronomical Journal}, 156\penalty0 (5):\penalty0 192,
  2018.

\bibitem[Quarles et~al.(2017)Quarles, Quintana, Lopez, Schlieder, and
  Barclay]{Quarles17}
Billy Quarles, Elisa~V Quintana, Eric~D Lopez, Joshua~E Schlieder, and Thomas
  Barclay.
\newblock Plausible compositions of the seven trappist-1 planets using
  long-term dynamical simulations.
\newblock \emph{arXiv preprint arXiv:1704.02261}, 2017.

\bibitem[Tamayo et~al.(2017)Tamayo, Rein, Petrovich, and Murray]{Tamayo17}
Daniel Tamayo, Hanno Rein, Cristobal Petrovich, and Norman Murray.
\newblock Convergent migration renders trappist-1 long-lived.
\newblock \emph{The Astrophysical Journal Letters}, 840\penalty0 (2):\penalty0
  L19, 2017.

\bibitem[Rivera et~al.(2010)Rivera, Laughlin, Butler, Vogt, Haghighipour, and
  Meschiari]{Rivera10}
Eugenio~J Rivera, Gregory Laughlin, R~Paul Butler, Steven~S Vogt, Nader
  Haghighipour, and Stefano Meschiari.
\newblock The lick-carnegie exoplanet survey: a uranus-mass fourth planet for
  gj 876 in an extrasolar laplace configuration.
\newblock \emph{The Astrophysical Journal}, 719\penalty0 (1):\penalty0 890,
  2010.

\bibitem[Jontof-Hutter et~al.(2014)Jontof-Hutter, Lissauer, Rowe, and
  Fabrycky]{Jontof14}
Daniel Jontof-Hutter, Jack~J Lissauer, Jason~F Rowe, and Daniel~C Fabrycky.
\newblock Kepler-79's low density planets.
\newblock \emph{The Astrophysical Journal}, 785\penalty0 (1):\penalty0 15,
  2014.

\bibitem[Buchhave et~al.(2016)Buchhave, Dressing, Dumusque, Rice, Vanderburg,
  Mortier, Lopez-Morales, Lopez, Lundkvist, Kjeldsen, et~al.]{Buchhave16}
Lars~A Buchhave, Courtney~D Dressing, Xavier Dumusque, Ken Rice, Andrew
  Vanderburg, Annelies Mortier, Mercedes Lopez-Morales, Eric Lopez, Mia~S
  Lundkvist, Hans Kjeldsen, et~al.
\newblock A 1.9 earth radius rocky planet and the discovery of a non-transiting
  planet in the kepler-20 system.
\newblock \emph{The Astronomical Journal}, 152\penalty0 (6):\penalty0 160,
  2016.

\bibitem[Hadden and Lithwick(2017)]{Hadden17}
Sam Hadden and Yoram Lithwick.
\newblock Kepler planet masses and eccentricities from ttv analysis.
\newblock \emph{The Astronomical Journal}, 154\penalty0 (1):\penalty0 5, 2017.

\bibitem[Grimm et~al.(2018)Grimm, Demory, Gillon, Dorn, Agol, Burdanov, Delrez,
  Sestovic, Triaud, Turbet, et~al.]{Grimm18}
Simon~L Grimm, Brice-Olivier Demory, Micha{\"e}l Gillon, Caroline Dorn, Eric
  Agol, Artem Burdanov, Laetitia Delrez, Marko Sestovic, Amaury~HMJ Triaud,
  Martin Turbet, et~al.
\newblock The nature of the trappist-1 exoplanets.
\newblock \emph{Astronomy \& Astrophysics}, 613:\penalty0 A68, 2018.

\bibitem[Wisdom(1980)]{Wisdom80}
Jack Wisdom.
\newblock The resonance overlap criterion and the onset of stochastic behavior
  in the restricted three-body problem.
\newblock \emph{The Astronomical Journal}, 85:\penalty0 1122--1133, 1980.

\bibitem[{Deck} et~al.(2013){Deck}, {Payne}, and {Holman}]{Deck13}
K.~M. {Deck}, M.~{Payne}, and M.~J. {Holman}.
\newblock {First-order Resonance Overlap and the Stability of Close Two-planet
  Systems}.
\newblock \emph{Astrophysical Journal}, 774:\penalty0 129, September 2013.
\newblock \doi{10.1088/0004-637X/774/2/129}.

\bibitem[Hadden and Lithwick(2018)]{Hadden18}
Sam Hadden and Yoram Lithwick.
\newblock A criterion for the onset of chaos in systems of two eccentric
  planets.
\newblock \emph{The Astronomical Journal}, 156\penalty0 (3):\penalty0 95, 2018.

\bibitem[{Chambers} et~al.(1996){Chambers}, {Wetherill}, and
  {Boss}]{Chambers96}
J.~E. {Chambers}, G.~W. {Wetherill}, and A.~P. {Boss}.
\newblock {The Stability of Multi-Planet Systems}.
\newblock \emph{Icarus}, 119:\penalty0 261--268, February 1996.
\newblock \doi{10.1006/icar.1996.0019}.

\bibitem[Quillen(2011)]{Quillen11}
Alice~C Quillen.
\newblock Three-body resonance overlap in closely spaced multiple-planet
  systems.
\newblock \emph{Monthly Notices of the Royal Astronomical Society},
  418\penalty0 (2):\penalty0 1043--1054, 2011.

\bibitem[Quillen and French(2014)]{Quillen14}
Alice~C Quillen and Robert~S French.
\newblock Resonant chains and three-body resonances in the closely packed inner
  uranian satellite system.
\newblock \emph{Monthly Notices of the Royal Astronomical Society},
  445\penalty0 (4):\penalty0 3959--3986, 2014.

\bibitem[Yoshinaga et~al.(1999)Yoshinaga, Kokubo, and Makino]{Yoshinaga99}
Keiko Yoshinaga, Eiichiro Kokubo, and Junichiro Makino.
\newblock The stability of protoplanet systems.
\newblock \emph{Icarus}, 139\penalty0 (2):\penalty0 328--335, 1999.

\bibitem[{Marzari} and {Weidenschilling}(2002)]{Marzari02}
F.~{Marzari} and S.~J. {Weidenschilling}.
\newblock {Eccentric Extrasolar Planets: The Jumping Jupiter Model}.
\newblock \emph{Icarus}, 156:\penalty0 570--579, April 2002.
\newblock \doi{10.1006/icar.2001.6786}.

\bibitem[Zhou et~al.(2007)Zhou, Lin, and Sun]{Zhou07}
Ji-Lin Zhou, Douglas~NC Lin, and Yi-Sui Sun.
\newblock Post-oligarchic evolution of protoplanetary embryos and the stability
  of planetary systems.
\newblock \emph{The Astrophysical Journal}, 666\penalty0 (1):\penalty0 423,
  2007.

\bibitem[Faber and Quillen(2007)]{Faber07}
Peter Faber and Alice~C Quillen.
\newblock The total number of giant planets in debris discs with central
  clearings.
\newblock \emph{Monthly Notices of the Royal Astronomical Society},
  382\penalty0 (4):\penalty0 1823--1828, 2007.

\bibitem[Smith and Lissauer(2009)]{Smith09}
Andrew~W Smith and Jack~J Lissauer.
\newblock Orbital stability of systems of closely-spaced planets.
\newblock \emph{Icarus}, 201\penalty0 (1):\penalty0 381--394, 2009.

\bibitem[{Matsumoto} et~al.(2012){Matsumoto}, {Nagasawa}, and
  {Ida}]{Matsumoto12}
Yuji {Matsumoto}, Makiko {Nagasawa}, and Shigeru {Ida}.
\newblock {The orbital stability of planets trapped in the first-order
  mean-motion resonances}.
\newblock \emph{Icarus}, 221\penalty0 (2):\penalty0 624--631, November 2012.
\newblock \doi{10.1016/j.icarus.2012.08.032}.

\bibitem[Pu and Wu(2015)]{Pu15}
Bonan Pu and Yanqin Wu.
\newblock Spacing of kepler planets: sculpting by dynamical instability.
\newblock \emph{The Astrophysical Journal}, 807\penalty0 (1):\penalty0 44,
  2015.

\bibitem[Obertas et~al.(2017)Obertas, Van~Laerhoven, and Tamayo]{Obertas17}
Alysa Obertas, Christa Van~Laerhoven, and Daniel Tamayo.
\newblock The stability of tightly-packed, evenly-spaced systems of earth-mass
  planets orbiting a sun-like star.
\newblock \emph{Icarus}, 293:\penalty0 52--58, 2017.
\newblock \doi{10.1016/j.icarus.2017.04.010}.

\bibitem[Rein and Tamayo(2015)]{ReinTamayo15}
Hanno Rein and Daniel Tamayo.
\newblock whfast: a fast and unbiased implementation of a symplectic
  wisdom--holman integrator for long-term gravitational simulations.
\newblock \emph{Monthly Notices of the Royal Astronomical Society},
  452\penalty0 (1):\penalty0 376--388, 2015.

\bibitem[Hussain and Tamayo(2019)]{Hussain19}
Naireen Hussain and Daniel Tamayo.
\newblock Fundamental limits from chaos on instability time predictions in
  compact planetary systems.
\newblock \emph{Monthly Notices of the Royal Astronomical Society},
  491\penalty0 (4):\penalty0 5258--5267, 2019.

\bibitem[Rice et~al.(2018)Rice, Rasio, and Steffen]{Rice18}
David~R Rice, Frederic~A Rasio, and Jason~H Steffen.
\newblock Survival of non-coplanar, closely packed planetary systems after a
  close encounter.
\newblock \emph{Monthly Notices of the Royal Astronomical Society},
  481\penalty0 (2):\penalty0 2205--2212, 2018.

\bibitem[Yalinewich and Petrovich(2019)]{Yalinewich19}
Almog Yalinewich and Cristobal Petrovich.
\newblock Nekhoroshev estimates for the survival time of tightly packed
  planetary systems.
\newblock \emph{arXiv preprint arXiv:1907.06660}, 2019.

\bibitem[Funk et~al.(2010)Funk, Wuchterl, Schwarz, Pilat-Lohinger, and
  Eggl]{Funk10}
B~Funk, G~Wuchterl, R~Schwarz, E~Pilat-Lohinger, and S~Eggl.
\newblock The stability of ultra-compact planetary systems.
\newblock \emph{Astronomy \& Astrophysics}, 516:\penalty0 A82, 2010.

\bibitem[Wu et~al.(2019)Wu, Zhang, Zhou, and Steffen]{Wu19}
Dong-Hong Wu, Rachel~C Zhang, Ji-Lin Zhou, and Jason~H Steffen.
\newblock Dynamical instability and its implications for planetary system
  architecture.
\newblock \emph{Monthly Notices of the Royal Astronomical Society},
  484\penalty0 (2):\penalty0 1538--1548, 2019.

\bibitem[{Murray} and {Dermott}(1999)]{Murray99}
C.~D. {Murray} and S.~F. {Dermott}.
\newblock \emph{{Solar System Dynamics}}.
\newblock Cambridge U. Press, Cambridge, 1999.

\bibitem[{Lithwick} and {Wu}(2011)]{Lithwick11secularchaos}
Y.~{Lithwick} and Y.~{Wu}.
\newblock {Theory of Secular Chaos and Mercury's Orbit}.
\newblock \emph{Astrophysical Journal}, 739:\penalty0 31, September 2011.
\newblock \doi{10.1088/0004-637X/739/1/31}.

\bibitem[{Batygin} et~al.(2015){Batygin}, {Morbidelli}, and
  {Holman}]{Batygin15c}
K.~{Batygin}, A.~{Morbidelli}, and M.~J. {Holman}.
\newblock {Chaotic Disintegration of the Inner Solar System}.
\newblock \emph{Astrophysical Journal}, 799:\penalty0 120, February 2015.
\newblock \doi{10.1088/0004-637X/799/2/120}.

\bibitem[Laplace(1784)]{Laplace84}
PS~Laplace.
\newblock M{\'e}moire sur les in{\'e}galit{\'e}s s{\'e}culaires des planetes et
  des satellites mem.
\newblock \emph{Acad. royale des Sciences de Paris, Oeuvres completes XI}, 49,
  1784.

\bibitem[Laskar(1990)]{Laskar90}
Jacques Laskar.
\newblock The chaotic motion of the solar system: A numerical estimate of the
  size of the chaotic zones.
\newblock \emph{Icarus}, 88\penalty0 (2):\penalty0 266--291, 1990.

\bibitem[Laskar(2000)]{Laskar00}
Jacques Laskar.
\newblock On the spacing of planetary systems.
\newblock \emph{Physical Review Letters}, 84\penalty0 (15):\penalty0 3240,
  2000.

\bibitem[Laskar and Petit(2017)]{Laskar17}
Jacques Laskar and AC~Petit.
\newblock Amd-stability and the classification of planetary systems.
\newblock \emph{Astronomy \& Astrophysics}, 605:\penalty0 A72, 2017.

\bibitem[Laskar and Gastineau(2009)]{Laskar09}
Jacques Laskar and Mickael Gastineau.
\newblock Existence of collisional trajectories of mercury, mars and venus with
  the earth.
\newblock \emph{Nature}, 459\penalty0 (7248):\penalty0 817--819, 2009.

\bibitem[Migaszewski et~al.(2012)Migaszewski, S{\l}onina, and
  Go{\'z}dziewski]{Migaszewski12}
Cezary Migaszewski, Mariusz S{\l}onina, and Krzysztof Go{\'z}dziewski.
\newblock A dynamical analysis of the kepler-11 planetary system.
\newblock \emph{Monthly Notices of the Royal Astronomical Society},
  427\penalty0 (1):\penalty0 770--789, 2012.

\bibitem[Petit et~al.(2017)Petit, Laskar, and Bou{\'e}]{Petit17}
Antoine~C Petit, Jacques Laskar, and Gwena{\"e}l Bou{\'e}.
\newblock Amd-stability in the presence of first-order mean motion resonances.
\newblock \emph{Astronomy \& Astrophysics}, 607:\penalty0 A35, 2017.

\bibitem[{Marzari}(2014)]{Marzari14}
F.~{Marzari}.
\newblock {Dynamical behaviour of multiplanet systems close to their stability
  limit}.
\newblock \emph{Monthly Notices of the Royal Astronomical Society},
  442:\penalty0 1110--1116, August 2014.
\newblock \doi{10.1093/mnras/stu929}.

\bibitem[Cincotta et~al.(2003)Cincotta, Giordano, and Sim{\'o}]{Cincotta03}
Pablo~Miguel Cincotta, Claudia~Marcela Giordano, and C~Sim{\'o}.
\newblock Phase space structure of multi-dimensional systems by means of the
  mean exponential growth factor of nearby orbits.
\newblock \emph{Physica D: Nonlinear Phenomena}, 182\penalty0 (3-4):\penalty0
  151--178, 2003.

\bibitem[{Wisdom} and {Holman}(1991)]{Wisdom91}
J.~{Wisdom} and M.~{Holman}.
\newblock {Symplectic maps for the n-body problem}.
\newblock \emph{Astronomical Journal}, 102:\penalty0 1528--1538, October 1991.
\newblock \doi{10.1086/115978}.

\bibitem[Volk and Gladman(2015)]{Volk15}
Kathryn Volk and Brett Gladman.
\newblock Consolidating and crushing exoplanets: Did it happen here?
\newblock \emph{The Astrophysical Journal Letters}, 806\penalty0 (2):\penalty0
  L26, 2015.

\bibitem[Tamayo et~al.(2016)Tamayo, Silburt, Valencia, Menou, Ali-Dib,
  Petrovich, Huang, Rein, van Laerhoven, Paradise, et~al.]{Tamayo16}
Daniel Tamayo, Ari Silburt, Diana Valencia, Kristen Menou, Mohamad Ali-Dib,
  Cristobal Petrovich, Chelsea~X Huang, Hanno Rein, Christa van Laerhoven, Adiv
  Paradise, et~al.
\newblock A machine learns to predict the stability of tightly packed planetary
  systems.
\newblock \emph{The Astrophysical Journal Letters}, 832\penalty0 (2):\penalty0
  L22, 2016.

\bibitem[Chen and Guestrin(2016)]{Chen16}
Tianqi Chen and Carlos Guestrin.
\newblock Xgboost: A scalable tree boosting system.
\newblock \emph{arXiv preprint arXiv:1603.02754}, 2016.

\bibitem[Lam and Kipping(2018)]{Lam18}
Christopher Lam and David Kipping.
\newblock A machine learns to predict the stability of circumbinary planets.
\newblock \emph{Monthly Notices of the Royal Astronomical Society},
  476\penalty0 (4):\penalty0 5692--5697, 2018.

\bibitem[Sessin and Ferraz-Mello(1984)]{Sessin84}
W~Sessin and S~Ferraz-Mello.
\newblock Motion of two planets with periods commensurable in the ratio 2:1
  solutions of the hori auxiliary system.
\newblock \emph{Celestial mechanics}, 32\penalty0 (4):\penalty0 307--332, 1984.

\bibitem[{Hadden}(2019)]{Hadden19}
Sam {Hadden}.
\newblock {An Integrable Model for the Dynamics of Planetary Mean-motion
  Resonances}.
\newblock \emph{Astronomical Journal}, 158\penalty0 (6):\penalty0 238, Dec
  2019.
\newblock \doi{10.3847/1538-3881/ab5287}.

\bibitem[Fabrycky et~al.(2014)Fabrycky, Lissauer, Ragozzine, Rowe, Steffen,
  Agol, Barclay, Batalha, Borucki, Ciardi, et~al.]{Fabrycky14}
Daniel~C Fabrycky, Jack~J Lissauer, Darin Ragozzine, Jason~F Rowe, Jason~H
  Steffen, Eric Agol, Thomas Barclay, Natalie Batalha, William Borucki, David~R
  Ciardi, et~al.
\newblock Architecture of kepler's multi-transiting systems. ii. new
  investigations with twice as many candidates.
\newblock \emph{The Astrophysical Journal}, 790\penalty0 (2):\penalty0 146,
  2014.

\bibitem[Weiss et~al.(2018)Weiss, Marcy, Petigura, Fulton, Howard, Winn,
  Isaacson, Morton, Hirsch, Sinukoff, et~al.]{Weiss18}
Lauren~M Weiss, Geoffrey~W Marcy, Erik~A Petigura, Benjamin~J Fulton, Andrew~W
  Howard, Joshua~N Winn, Howard~T Isaacson, Timothy~D Morton, Lea~A Hirsch,
  Evan~J Sinukoff, et~al.
\newblock The california-kepler survey. v. peas in a pod: planets in a kepler
  multi-planet system are similar in size and regularly spaced.
\newblock \emph{The Astronomical Journal}, 155\penalty0 (1):\penalty0 48, 2018.

\bibitem[Chirikov(1979)]{Chirikov79}
Boris~V Chirikov.
\newblock A universal instability of many-dimensional oscillator systems.
\newblock \emph{Physics reports}, 52\penalty0 (5):\penalty0 263--379, 1979.

\bibitem[Bergstra et~al.(2013)Bergstra, Yamins, and Cox]{Bergstra13}
James Bergstra, Daniel Yamins, and David~Daniel Cox.
\newblock Making a science of model search: Hyperparameter optimization in
  hundreds of dimensions for vision architectures.
\newblock 2013.

\bibitem[{Petit} et~al.(2018){Petit}, {Laskar}, and {Bou{\'e}}]{Petit18}
Antoine~C. {Petit}, Jacques {Laskar}, and Gwena{\"e}l {Bou{\'e}}.
\newblock {Hill stability in the AMD framework}.
\newblock \emph{Astronomy \& Astrophysics}, 617:\penalty0 A93, Sep 2018.
\newblock \doi{10.1051/0004-6361/201833088}.

\bibitem[Batygin et~al.(2015)Batygin, Deck, and Holman]{Batygin15}
Konstantin Batygin, Katherine~M Deck, and Matthew~J Holman.
\newblock Dynamical evolution of multi-resonant systems: The case of gj 876.
\newblock \emph{The Astronomical Journal}, 149\penalty0 (5):\penalty0 167,
  2015.

\bibitem[Chen and Kipping(2016)]{ChenKip16}
Jingjing Chen and David Kipping.
\newblock Probabilistic forecasting of the masses and radii of other worlds.
\newblock \emph{The Astrophysical Journal}, 834\penalty0 (1):\penalty0 17,
  2016.

\bibitem[{Chatterjee} et~al.(2008){Chatterjee}, {Ford}, {Matsumura}, and
  {Rasio}]{Chatterjee08}
S.~{Chatterjee}, E.~B. {Ford}, S.~{Matsumura}, and F.~A. {Rasio}.
\newblock {Dynamical Outcomes of Planet-Planet Scattering}.
\newblock \emph{Astrophysical Journal}, 686:\penalty0 580--602, October 2008.
\newblock \doi{10.1086/590227}.

\bibitem[{Juri{\'c}} and {Tremaine}(2008)]{Juric08}
M.~{Juri{\'c}} and S.~{Tremaine}.
\newblock {Dynamical Origin of Extrasolar Planet Eccentricity Distribution}.
\newblock \emph{Astrophysical Journal}, 686:\penalty0 603--620, October 2008.
\newblock \doi{10.1086/590047}.

\bibitem[Butler et~al.(2006)Butler, Wright, Marcy, Fischer, Vogt, Tinney,
  Jones, Carter, Johnson, McCarthy, et~al.]{Butler06}
R~Paul Butler, JT~Wright, GW~Marcy, DA~Fischer, SS~Vogt, CG~Tinney, HRA Jones,
  BD~Carter, JA~Johnson, C~McCarthy, et~al.
\newblock Catalog of nearby exoplanets.
\newblock \emph{The Astrophysical Journal}, 646\penalty0 (1):\penalty0 505,
  2006.

\bibitem[Van~Eylen and Albrecht(2015)]{vanEylen15}
Vincent Van~Eylen and Simon Albrecht.
\newblock Eccentricity from transit photometry: small planets in kepler
  multi-planet systems have low eccentricities.
\newblock \emph{The Astrophysical Journal}, 808\penalty0 (2):\penalty0 126,
  2015.

\bibitem[Xie et~al.(2016)Xie, Dong, Zhu, Huber, Zheng, De~Cat, Fu, Liu, Luo,
  Wu, et~al.]{Xie16}
Ji-Wei Xie, Subo Dong, Zhaohuan Zhu, Daniel Huber, Zheng Zheng, Peter De~Cat,
  Jianning Fu, Hui-Gen Liu, Ali Luo, Yue Wu, et~al.
\newblock Exoplanet orbital eccentricities derived from lamost--kepler
  analysis.
\newblock \emph{Proceedings of the National Academy of Sciences}, 113\penalty0
  (41):\penalty0 11431--11435, 2016.

\bibitem[Burke et~al.(2015)Burke, Christiansen, Mullally, Seader, Huber, Rowe,
  Coughlin, Thompson, Catanzarite, Clarke, et~al.]{Burke15}
Christopher~J Burke, Jessie~L Christiansen, F~Mullally, Shawn Seader, Daniel
  Huber, Jason~F Rowe, Jeffrey~L Coughlin, Susan~E Thompson, Joseph
  Catanzarite, Bruce~D Clarke, et~al.
\newblock Terrestrial planet occurrence rates for the kepler gk dwarf sample.
\newblock \emph{The Astrophysical Journal}, 809\penalty0 (1):\penalty0 8, 2015.

\bibitem[Holman and Murray(2005)]{Holman05}
Matthew~J Holman and Norman~W Murray.
\newblock The use of transit timing to detect terrestrial-mass extrasolar
  planets.
\newblock \emph{Science}, 307\penalty0 (5713):\penalty0 1288--1291, 2005.

\bibitem[{Hadden} et~al.(2019){Hadden}, {Barclay}, {Payne}, and
  {Holman}]{Hadden2019_tess}
Sam {Hadden}, Thomas {Barclay}, Matthew~J. {Payne}, and Matthew~J. {Holman}.
\newblock {Prospects for TTV Detection and Dynamical Constraints with TESS}.
\newblock \emph{The Astrophysical Journal}, 158\penalty0 (4):\penalty0 146,
  October 2019.
\newblock \doi{10.3847/1538-3881/ab384c}.

\bibitem[Lithwick et~al.(2012)Lithwick, Xie, and Wu]{Lithwick12}
Yoram Lithwick, Jiwei Xie, and Yanqin Wu.
\newblock Extracting planet mass and eccentricity from ttv data.
\newblock \emph{The Astrophysical Journal}, 761\penalty0 (2):\penalty0 122,
  2012.

\bibitem[Jontof-Hutter et~al.(2016)Jontof-Hutter, Ford, Rowe, Lissauer,
  Fabrycky, Van~Laerhoven, Agol, Deck, Holczer, and Mazeh]{Jontof16}
Daniel Jontof-Hutter, Eric~B Ford, Jason~F Rowe, Jack~J Lissauer, Daniel~C
  Fabrycky, Christa Van~Laerhoven, Eric Agol, Katherine~M Deck, Tomer Holczer,
  and Tsevi Mazeh.
\newblock Secure mass measurements from transit timing: 10 kepler exoplanets
  between 3 and 8 mearth with diverse densities and incident fluxes.
\newblock \emph{The Astrophysical Journal}, 820\penalty0 (1):\penalty0 39,
  2016.

\bibitem[Volk and Malhotra(2020)]{Volk20}
Kathryn Volk and Renu Malhotra.
\newblock Dynamical instabilities in systems of multiple short-period planets
  are likely driven by secular chaos: a case study of kepler-102.
\newblock \emph{arXiv preprint arXiv:2003.05040}, 2020.

\bibitem[Barnes and Quinn(2004)]{Barnes04}
Rory Barnes and Thomas Quinn.
\newblock The (in) stability of planetary systems.
\newblock \emph{The Astrophysical Journal}, 611\penalty0 (1):\penalty0 494,
  2004.

\bibitem[{Tremaine}(2015)]{Tremaine15}
Scott {Tremaine}.
\newblock {The Statistical Mechanics of Planet Orbits}.
\newblock \emph{Astrophysical Journal}, 807\penalty0 (2):\penalty0 157, Jul
  2015.
\newblock \doi{10.1088/0004-637X/807/2/157}.

\bibitem[Izidoro et~al.(2017)Izidoro, Ogihara, Raymond, Morbidelli, Pierens,
  Bitsch, Cossou, and Hersant]{Izidoro17}
Andre Izidoro, Masahiro Ogihara, Sean~N Raymond, Alessandro Morbidelli, Arnaud
  Pierens, Bertram Bitsch, Christophe Cossou, and Franck Hersant.
\newblock Breaking the chains: hot super-earth systems from migration and
  disruption of compact resonant chains.
\newblock \emph{Monthly Notices of the Royal Astronomical Society},
  470\penalty0 (2):\penalty0 1750--1770, 2017.

\bibitem[Murray and Holman(1997)]{Murray97}
N~Murray and M~Holman.
\newblock Diffusive chaos in the outer asteroid belt.
\newblock \emph{The Astronomical Journal}, 114:\penalty0 1246--1259, 1997.

\bibitem[{Gladman}(1993)]{Gladman93}
B.~{Gladman}.
\newblock {Dynamics of systems of two close planets}.
\newblock \emph{Icarus}, 106:\penalty0 247, November 1993.
\newblock \doi{10.1006/icar.1993.1169}.

\bibitem[Hadden and Lithwick(2016)]{Hadden16}
Sam Hadden and Yoram Lithwick.
\newblock Numerical and analytical modeling of transit timing variations.
\newblock \emph{The Astrophysical Journal}, 828\penalty0 (1):\penalty0 44,
  2016.

\bibitem[Rein and Liu(2012)]{Rein12}
Hanno Rein and S-F Liu.
\newblock Rebound: an open-source multi-purpose n-body code for collisional
  dynamics.
\newblock \emph{Astronomy \& Astrophysics}, 537:\penalty0 A128, 2012.

\bibitem[Rein and Tamayo(2017)]{ReinTamayo17}
Hanno Rein and Daniel Tamayo.
\newblock A new paradigm for reproducing and analysing n-body simulations of
  planetary systems.
\newblock \emph{Monthly Notices of the Royal Astronomical Society},
  467\penalty0 (2):\penalty0 2377, 2017.
\newblock \doi{10.1093/mnras/stx232}.
\newblock URL \url{+ http://dx.doi.org/10.1093/mnras/stx232}.

\end{thebibliography}

\end{document}